# On the User Selection for MIMO Broadcast Channels

Alireza Bayesteh and Amir K. Khandani

Electrical and Computer Engineering Department

University of Waterloo, Waterloo, ON, Canada

Email:{alireza,khandani}@cst.uwaterloo.ca

# On the User Selection for MIMO Broadcast Channels

Alireza Bayesteh, and Amir K. Khandani

Dept. of Electrical Engineering

University of Waterloo

Waterloo, ON, N2L 3G1

alireza, khandani@shannon2.uwaterloo.ca

**Abstract**

In this paper, a downlink communication system, in which a Base Station (BS) equipped with $M$ antennas communicates with $N$ users each equipped with $K$ receive antennas, is considered. An efficient suboptimum algorithm is proposed for selecting a set of users in order to maximize the sum-rate throughput of the system. For the asymptotic case when $N$ tends to infinity, the necessary and sufficient conditions in order to achieve the maximum sum-rate throughput, such that the difference between the achievable sum-rate and the maximum value approaches zero, is derived. The complexity of our algorithm is investigated in terms of the required amount of feedback from the users to the base station, as well as the number of searches required for selecting the users. It is shown that the proposed method is capable of achieving a large portion of the sum-rate capacity, with a very low complexity.

Financial supports provided by Nortel, and the corresponding matching funds by the Federal government: Natural Sciences and Engineering Research Council of Canada (NSERC) and Province of Ontario: Ontario Centres of Excellence (OCE) are gratefully acknowledged.

## I. INTRODUCTION

Multiple-input multiple-output (MIMO) systems have proved their ability to achieve high bit rates on a scattering wireless network [1]. In a MIMO broadcast channel, the base station equipped with multiple antennas communicates with several multiple-antenna users. Recently, there has been a lot of interest in characterizing the capacity region of this channel [2], [3], [4], [5]. In [2]- [4], it has been shown that the sum-rate capacity of MIMO broadcast channels can be achieved by applying dirty-paper coding (DPC) [6] at the transmitter. Practical schemes for approximate implementation of DPC are proposed in [7], [8], [9], [10], [11], [12]. However, achieving the theoretical limits promised by DPC faces many challenges.

In a network with a large number of users, the base station can increase the throughput by selecting the best set of users to communicate with. This results in the so-called "multiuser diversity" gain [13], [14]. However, achieving the optimum multiuser diversity gain requires an exhaustive search over all possible combination of the users, which is not practical for large-scale networks. To overcome this problem, references [15] and [16] propose sub-optimum methods for user selection. These methods exploit the multiuser diversity gain, but are based on assuming DPC at the base station.

To avoid the complexity of DPC, the simple precoding scheme of "zero-forcing beam-forming", which is also called "channel inversion", is considered by some authors [17], [18], [19], [20]. In these works, it is assumed that the users are equipped with a single antenna. Using zero-forcing beam-forming, the downlink channel with $M$ transmit antennas is decomposed into $N \leq M$ interference-free subchannels, serving $N$ users. Unfortunately, in cases that the number of users is equal to the number of transmit antennas, this method does not offer a good performance [20]. However, the case of $N > M$ is more common in practical networks. In this case, selecting the best set of users improves the performance of this scheme significantly [5] , [21] (multiuser diversity gain). Due to the high complexity of selecting the best set, reference [22] proposes a suboptimum algorithm for user selection in order to maximize the sum-rate. This algorithm is based on using zero-forcing beam-forming at the transmitter. The complexity of this algorithm is shown to be $O(M^3N)$.

To achieve a good performance by using zero-forcing beam-forming, the selected sub-channels must have high gains and be nearly orthogonal to each other. As the number of users increases, it becomes easier to satisfy these requirements. However, the exhaustive search for selecting the best set of users is very complex. In [23], the authors propose a suboptimum algorithm for selecting such a set of users in a downlink environment with large number of single-antenna users. This algorithm is similar to the greedy algorithm proposed in [15], with the difference in using an orthogonality threshold for selecting the users in each step. As a result, the channel vectors of the selected users become nearly orthogonal to each other with considerable gains. It has been shown that using this algorithm, the optimum sum-rate throughput of the system is asymptotically achieved as $N \to \infty$. However, in their approach, the base station must have perfect Channel State Information (CSI) for all users.

To avoid the huge amount of feedback required by providing perfect CSI to the base station, reference [24] proposes a downlink transmission scheme based on random beam-forming relying on partial CSI at the transmitter. In this scheme, the base station randomly constructs $M$ orthogonal beams and transmits data to the users with the maximum Signal to Interference plus Noise Ratio (SINR) for each beam. Therefore, only the value of maximum SINR, and the index of the beam for which the maximum SINR is achieved, are fed back to the base station for each user. This significantly reduces the amount of feedback. Reference [24] shows that when the number of users tends to infinity, the optimum sum-rate throughput can be achieved. However, for practical number of users, it does not perform well [23].

In this paper, we consider a MIMO-BC with large number of users and propose an efficient sub-optimum algorithm that assigns the coordinates of transmission space to different users in order to achieve the best performance in terms of the sum-rate throughput. It is assumed that the zero-forcing beam-forming is used at the base station as the precoding scheme. The algorithm starts by setting a threshold value. By applying Singular Value Decomposition (SVD) to all users' channel matrices, only the eigenvectors whose corresponding singular values are above the set threshold are considered. Then, among these candidate eigenvectors, the algorithm chooses a set of size $M$ which are nearly orthogonal to each other. For the asymptotic case of $N \to \infty$, we give the necessary and sufficient conditions for the threshold value in order to achieve the optimum sum-rate capacity, such that the difference between the sum-rates approaches zero. The proposed algorithm follows the same approach as that of [23], with a difference in the user selection strategy. The main advantage of our algorithm is that the coordinates are selected among the eigenvectors with singular values above a given threshold, and for the rest of the eigenvectors no information is sent to the base station. Therefore, the complexity of search and the amount of feedback required at the base station is significantly reduced. Indeed, we give the necessary and sufficient conditions for the threshold value in order to achieve the optimum sum-rate, such that the difference between the achievable sum-rate and the optimum value approaches zero.

This paper is organized as follows. In section II, we introduce the system model, and describe the proposed algorithm. Sections III and IV are devoted to analyzing the performance, in terms of the sum-rate throughput, and the complexity of our proposed algorithm, respectively. Finally, section V concludes the paper.

Throughout this paper, the norm of the vectors are denoted by $\|.\|$, the Hermitian operation is denoted by $(.)^*$, and the determinant and the trace operations are denoted by $\det(.)$ and $\mathrm{Tr}(.)$, respectively. $\mathbb{E}\{.\}$ represents the expectation, notation "ln" is used for the natural logarithm, and the rates are expressed in *nats*. $\mathrm{RH}(.)$ represents the right hand side of the equations. For any given functions $f(N)$ and $g(N)$, $f(N) = O(g(N))$ is equivalent to $\lim_{N\to\infty} \left|\frac{f(N)}{g(N)}\right| < \infty$, $f(N) = o(g(N))$ is equivalent to $\lim_{N\to\infty} \left|\frac{f(N)}{g(N)}\right| = 0$, $f(N) = \Omega(g(N))$ is equivalent to $\lim_{N\to\infty} \frac{f(N)}{g(N)} > 0$, $f(N) = \omega(g(N))$ is equivalent to $\lim_{N\to\infty} \frac{f(N)}{g(N)} = \infty$, and $f(N) = \Theta(g(N))$ is equivalent to $\lim_{N\to\infty} \frac{f(N)}{g(N)} = c$, where $0 < c < \infty$.

## II. System Model

In this work, a MIMO-BC in which a base station equipped with $M$ antennas communicates with $N$ users, each equipped with $K$ antennas, is considered. The channel between each user and the base station is modeled as a zero-mean circularly symmetric Gaussian matrix (Rayleigh fading). The received vector by user *k* can be written as

$$\boldsymbol{y}_k = \boldsymbol{H}_k \boldsymbol{x} + \boldsymbol{n}_k, \tag{1}$$

where $\boldsymbol{x} \in \mathbb{C}^{M\times 1}$ is the transmitted signal, $\boldsymbol{H}_k \in \mathbb{C}^{K\times M}$ is the channel matrix from the transmitter to the $k$th user (assumed to be known at the receiver side), and $\boldsymbol{n}_k \in \mathbb{C}^{K\times 1} \sim \mathcal{CN}(\mathbf{0}, \boldsymbol{I}_K)$ is the noise vector at this receiver. We assume that the transmitter has an average power constraint $P$, i.e. $\mathbb{E}\left\{\mathrm{Tr}(\boldsymbol{x}\boldsymbol{x}^*)\right\} \leq P$. We consider a block fading model in which each $\boldsymbol{H}_k$ is constant for the duration of a frame. The frame itself is assumed to be long enough to allow communication at rates close to the capacity.

The maximum achievable sum-rate capacity in MIMO-BC, denoted as $\mathcal{R}_{\text{Opt}}$, is equal to [2]

$$\mathcal{R}_{\text{Opt}} = \mathbb{E}\left\{ \max_{\substack{\boldsymbol{Q}_n \\ \sum \text{Tr}(\boldsymbol{Q}_\text{n})=\text{P}}} \log\det\left(\boldsymbol{I}_M + \sum_{n=1}^{N} \boldsymbol{H}_n^* \boldsymbol{Q}_n \boldsymbol{H}_n\right)\right\}, \tag{2}$$

where $\boldsymbol{Q}_n$ is the transmit covariance matrix of the $n$th user, and the expectation is taken over the channel matrices $\boldsymbol{H}_1, \cdots, \boldsymbol{H}_N$. The capacity achieving transmission strategy is shown to involve at least $M$, and at most $M^2$ data streams in total [25]. However, experimental results show that $M$ data streams are adequate to achieve a significant portion of the capacity [15], [16].

As discussed earlier, the capacity achieving strategy in a downlink environment requires applying dirty-paper coding at the base station, which is not practical in many applications. For this reason, it is desirable to utilize a precoding scheme with less complexity. Among the known precoding schemes, zero-forcing beam-forming has received considerable attention, as it uses a simple structure of channel matrix inversion. This scheme results in having $M$ interference-free sub-channels. Although this scheme does not yield a good performance for the case $M = N$ [20][1], for the case of $N > M$, which is more common in wireless networks, by selecting an appropriate set of dimensions, the corresponding performance is shown to be good [23], [22], [26]. In this work, using zero-forcing beam-forming at the base station, we propose an efficient algorithm to find $M$ coordinates for data transmission, focusing on maximizing the sum-rate throughput.

## III. Proposed Algorithm

As mentioned earlier, to maximize the sum-rate using zero-forcing beam-forming, the selected eigenvectors must be nearly orthogonal to each other, and their corresponding singular values be sufficiently large. The measure of orthogonality between two $M \times 1$ vectors $\boldsymbol{\upsilon}$ and $\boldsymbol{\psi}$ is defined as,

$$z(\boldsymbol{\upsilon}, \boldsymbol{\psi}) = \frac{|\boldsymbol{\upsilon}^* \boldsymbol{\psi}|^2}{\|\boldsymbol{\upsilon}\|^2 \|\boldsymbol{\psi}\|^2}. \tag{3}$$

[1]The result is derived for the case of single-antenna users

It is evident that the smaller is $z(\boldsymbol{\upsilon}, \boldsymbol{\psi})$, the more orthogonal will be $\boldsymbol{\upsilon}$ and $\boldsymbol{\psi}$.

Using Singular Value Decomposition (SVD), $\boldsymbol{H}_k$ can be written as

$$\boldsymbol{H}_k = \boldsymbol{U}_k \boldsymbol{\Lambda}_k \boldsymbol{V}_k^*, \tag{4}$$

where $\boldsymbol{\Lambda}_k$ is a $K \times M$ diagonal matrix containing the singular values of $\boldsymbol{H}_k$, $\boldsymbol{U}_k$ and $\boldsymbol{V}_k$ are $K \times K$ and $M \times M$ unitary matrices, respectively. Multiplying both sides of (1) by $\boldsymbol{U}_{k,j}^*$, where $\boldsymbol{U}_{k,j}$ is the $j$th column of $\boldsymbol{U}_k$, it is easy to show that

$$r_{k,j} = \boldsymbol{g}_{k,j} \boldsymbol{x} + w_{k,j}. \tag{5}$$

In the above equation, $r_{k,j} = \boldsymbol{U}_{k,j}^* \boldsymbol{y}_k$, $\boldsymbol{g}_{k,j} = \sqrt{\lambda_j(k)} \boldsymbol{V}_{k,j}^*$, where $\boldsymbol{V}_{k,j}$ is the $j$th column of $\boldsymbol{V}_k$ and $\sqrt{\lambda_j(k)}$ is the $j$th singular value of $\boldsymbol{H}_k$, corresponding to $\boldsymbol{V}_{k,j}$, and $w_{k,j} \sim \mathcal{CN}(0,1)$ is AWGN. This equation suggests that for selecting the dimensions with high gains, the norm of the equivalent channel introduced by (5), namely $\boldsymbol{g}_{k,j}$, which is equal to $\sqrt{\lambda_j(k)}$, can be compared with a threshold. This threshold is set by the base station at the beginning of the transmission. Using such a threshold reduces the amount of feedback and the size of search space for selecting the coordinates. To satisfy the orthogonality criterion, the base station can perform an exhaustive search for finding the "most orthogonal set"[2] among the pre-selected eigenvectors. Due to the large complexity of exhaustive search, the coordinates can be chosen one by one. In other words, in each step the eigenvector which is the most orthogonal to the

[2]In general, the orthogonality of a set $\{\boldsymbol{h}_i\}_{i=1}^M$ can be measured by the orthogonality defect, defined as $\frac{\prod_{i=1}^M \|\boldsymbol{h}_i\|^2}{\det(\boldsymbol{H}\boldsymbol{H}^*)}$, where $\boldsymbol{H} = [\boldsymbol{h}_1^T | \cdots | \boldsymbol{h}_M^T]^T$.

previously selected coordinates, is selected. The first coordinate is chosen as the eigenvector with the maximum corresponding singular value. The steps of the algorithm are given in the following:

**Proposed Algorithm (Algorithm 1):**

1. Using SVD, each user computes the eigenvectors and singular values of its channel matrix and sends back the singular values which are larger than a predetermined threshold $t$, along with their corresponding “right” eigenvectors, to the base station. The indices of these eigenvectors form the following set:

$$\mathcal{S}_0 = \{(k,j) | \quad \lambda_j(k) > t\}. \tag{6}$$

2. Base station selects the index in $\mathcal{S}_0$, corresponding to the maximum eigenvalue. Let us define this index as $(s_1, d_1)$, i.e., the $d_1$th eigenvector of the $s_1$th user.

3. Define

$$\mathcal{S}_1 \quad = \quad \mathcal{S}_0 - \{(s_1, d_1)\},$$

and

$$\gamma_{k,j}(1) \quad = \quad z(\boldsymbol{V}_{s_1,d_1}, \boldsymbol{V}_{k,j}), \quad \forall (k,j) \in \mathcal{S}_1, \tag{7}$$

where $z(.,.)$ is defined in (3). Note that as $\|\boldsymbol{V}_{k,j}\| = \|\boldsymbol{V}_{s_1,d_1}\| = 1$, $z(\boldsymbol{V}_{s_1,d_1}, \boldsymbol{V}_{k,j}) = |\boldsymbol{V}^*_{s_1,d_1}\boldsymbol{V}_{k,j}|^2$.

4. For $\quad 2 \leq m \leq M$, repeat the followings:

$$\begin{aligned} (s_m, d_m) &= \arg \min_{(k,j)\in\mathcal{S}_{m-1}} \gamma_{k,j}(m-1) \\ \mathcal{S}_m &= \mathcal{S}_{m-1} - \{(s_m, d_m)\} \\ \gamma_{k,j}(m) &= z(\boldsymbol{V}_{s_m,d_m}, \boldsymbol{V}_{k,j}) + \gamma_{k,j}(m-1), \qquad \forall (k,j) \in \mathcal{S}_m. \end{aligned} \tag{8}$$

In the above, $\gamma_{k,j}(m-1) = \sum_{i=1}^{m-1} z(\boldsymbol{V}_{s_i,d_i}, \boldsymbol{V}_{k,j})$ is used as the measure of orthogonality between a candidate eigenvector $\boldsymbol{V}_{k,j}$ and the set of previously selected eigenvectors, $\{\boldsymbol{V}_{s_i,d_i}\}_{i=1}^{m-1}$. Since these eigenvectors are nearly orthogonal to each other by the algorithm, with a good approximation, $\gamma_{k,j}(m-1)$ can be interpreted as the square magnitude of the projection of $\boldsymbol{V}_{k,j}$ over the sub-space spanned by $\{\boldsymbol{V}_{s_i,d_i}\}_{i=1}^{m-1}$. It is obvious that the smaller is this projection, the more orthogonal will be $\boldsymbol{V}_{k,j}$ to this sub-space. The recursive structure of $\gamma_{k,j}(m)$ facilitates its computation at each step of the algorithm.

After selecting the dimensions, we construct the "selected coordinate matrix" as

$$\boldsymbol{\mathcal{H}} = \left[\boldsymbol{g}_{s_1,d_1}^T \,|\, \boldsymbol{g}_{s_2,d_2}^T \,|\, \cdots \,|\, \boldsymbol{g}_{s_M,d_M}^T\right]^T. \tag{9}$$

Using zero-forcing beam-forming, the transmitted vector $\boldsymbol{x}$ can be written as

$$\boldsymbol{x} = \boldsymbol{\mathcal{H}}^{-1}\boldsymbol{u}, \tag{10}$$

where $\boldsymbol{u} = [u_{s_1,d_1}, \cdots, u_{s_M,d_M}]^T$ is the information vector. Using (5) and (10), the received signal over the $m$th coordinate is equal to

$$\begin{aligned} r_{s_m,d_m} &= \boldsymbol{U}_{s_m,d_m}^* \boldsymbol{y}_{s_m} \\ &= \boldsymbol{g}_{s_m,d_m}\boldsymbol{x} + w_{s_m,d_m} \\ &= \boldsymbol{g}_{s_m,d_m}\boldsymbol{\mathcal{H}}^{-1}\boldsymbol{u} + w_{s_m,d_m} \\ &= u_{s_m,d_m} + w_{s_m,d_m}. \end{aligned} \tag{11}$$

It can be seen that by applying zero-forcing beam-forming, the downlink channel is decomposed to $M$ interference-free sub-channels.

## IV. Performance Analysis

In this section, we examine the performance of our proposed algorithm in terms of the sum-rate throughput. First, we consider the asymptotic case of $N \to \infty$.

### A. Asymptotic Analysis

The sum-rate capacity of MIMO-BC has been shown to scale as $M \log \log N$, as $N$ tends to infinity [24]. This implies that to achieve the optimum sum-rate, the singular values corresponding to the selected dimensions must behave like $\log N$. In other words, the threshold value should scale as $\log N$. The following theorems indicates this fact with more details:

**Theorem 1** *The necessary condition to achieve* $\lim_{N\to\infty} \mathcal{R}_{\text{Opt}} - \mathcal{R}_{\text{Prop}} = 0$ *is having*

$$t = \log N + (M + K - 2) \log \log N - \rho(N), \tag{12}$$

*where* $\rho(N)$ *satisfies*

$$\rho(N) \quad \sim \quad o(\log N),$$

*and*

$$\rho(N) \quad \sim \quad \log \log \log \log N + \log[\Gamma(K)\Gamma(M)] + \omega \left( \frac{1}{\log \log \log N} \right).$$

**Proof -** We show that by violating any of the above conditions, the optimum sum-rate can not be achieved.

*The necessity of* $\rho(N) \sim o(\log N)$*:*

It is sufficient to show that

$$\lim_{N\to\infty} \frac{t}{\log N} = 1. \tag{13}$$

For this purpose, we consider the following cases:

***Case I;*** $\lim_{N\to\infty} t = \infty, \lim_{N\to\infty} \dfrac{t}{\log N} < 1$**:** The achievable sum-rate of the proposed method, denoted by $\mathcal{R}_{\text{Prop}}$, can be upper-bounded as

$$\begin{aligned}\mathcal{R}_{\text{Prop}} &\leq \mathbb{E}\left\{\max_{\substack{P_i \\ \sum_{i=1}^M P_i = P}} \sum_{i=1}^{M} \log(1 + P_i \|\boldsymbol{g}_{s_i,d_i}\|^2)\right\} \\ &= \mathbb{E}\left\{\max_{\substack{P_i \\ \sum_{i=1}^M P_i = P}} \sum_{i=1}^{M} \log(1 + P_i \lambda_{d_i}(s_i))\right\}, \end{aligned} \tag{14}$$

where $\boldsymbol{g}_{s_i,d_i}$ and $\lambda_{d_i}(s_i)$ are defined in (5).

Since the optimum sum-rate is shown to be $M \log\left(\frac{P}{M}\log N + O(\log\log N)\right)$ [24], we have

$$\begin{aligned}\mathcal{R}_{\text{Opt}} - \mathcal{R}_{\text{Prop}} &\geq M\log\left(\frac{P}{M}\log N + O(\log\log N)\right) - \mathbb{E}\left\{\max_{\substack{P_i \\ \sum_{i=1}^M P_i = P}} \sum_{i=1}^{M} \log(1 + P_i \lambda_{d_i}(s_i))\right\}, \\ &= M\log\left(\frac{P}{M}\log N + O(\log\log N)\right) - \\ &\quad \mathbb{E}\left\{\max_{\substack{P_i \\ \sum_{i=1}^M P_i = P}} \sum_{i=1}^{M} \log(P_i \lambda_{d_i}(s_i)) + \log\left(1 + \frac{1}{P_i \lambda_{d_i}(s_i)}\right)\right\}. \end{aligned} \tag{15}$$

The right hand side of the above equation can be written as follows:

$$\begin{aligned}\text{RH}(15) &\overset{(a)}{\geq} \min_{\substack{P_i \\ \sum_{i=1}^M P_i = P}} \log\left(\frac{(P/M)^M}{\prod_{i=1}^M P_i}\right) + M\log\left(\log N + O(\log\log N)\right) \\ &\quad - \sum_{i=1}^{M} \mathbb{E}\left\{\log \lambda_{d_i}(s_i)\right\} + O\left(\frac{1}{t}\right) \\ &= M\log\left(\log N + O(\log\log N)\right) - \sum_{i=1}^{M} \mathbb{E}\left\{\log \lambda_{d_i}(s_i)\right\} + O\left(\frac{1}{t}\right) \\ &\overset{(b)}{\geq} M\log\left(\log N + O(\log\log N)\right) - \mathbb{E}\left\{\max_{k=1,\cdots,N} \log \lambda_{\max}(\boldsymbol{H}_k)\right\} \\ &\quad -(M-1)\mathbb{E}\left\{\log \lambda / \lambda > t\right\} + O\left(\frac{1}{t}\right), \end{aligned} \tag{16}$$

where $\lambda_{\max}(\boldsymbol{A})$ is the maximum singular value of $\boldsymbol{A}\boldsymbol{A}^*$, and $\lambda$ is a random variable, denoting an unordered eigenvalue of a $K \times K$ Wishart matrix. $(a)$ comes from using the approximation $\log(1+x) \sim O(x)$, $x \ll 1$, noting that the solution to the maximization problem (14) satisfies $P_i \lambda_{s_i}(d_i) \gg 1$, $i = 1, \cdots, M$. $(b)$ results from the fact that excluding the largest maximum singular value from the set of singular values, which are greater than $t$, reduces the expectation in the second line of (16). In writing $(b)$, we also used the fact that the eigenvectors and their corresponding singular values of a circularly symmetric Gaussian matrix are independent. The distribution of $\lambda$, denoting as $f(\lambda)$ is derived in [1] as

$$f(\lambda) = \frac{1}{K} \sum_{i=0}^{K-1} \frac{i!}{(M-K+i)!} [L_i^{M-K}(\lambda)]^2 \lambda^{M-K} \exp(-\lambda), \tag{17}$$

where $L_i^{M-K}(\lambda)$ is the associated Laguerre polynomial of order $k$ [27]. Using the above equation, it is easy to show that

$$\begin{aligned} \mathbb{E}\left\{\log \lambda / \lambda > t\right\} &= \frac{\int_t^\infty \log \lambda f(\lambda) d\lambda}{1 - F(t)} \\ &= \log t + \frac{\int_t^\infty \frac{1-F(\lambda)}{\lambda} d\lambda}{1 - F(t)} \\ &\sim \log t + O\left(\frac{1}{t}\right). \end{aligned} \tag{18}$$

Indeed, we can write

$$\mathbb{E}\left\{\max_{k=1,\cdots,N} \log \lambda_{\max}(\boldsymbol{H}_k)\right\} \leq \mathbb{E}\left\{\max_{k=1,\cdots,N} \log \|\boldsymbol{H}_k\|^2\right\}, \tag{19}$$

where $\|\boldsymbol{A}\|^2$ denotes the Frobenius norm of matrix $\boldsymbol{A}$. In [24], it has been shown that with probability one,

$$\max_{k=1,\cdots,N} \|\boldsymbol{H}_k\|^2 \sim \log N + O(\log\log N).$$

Therefore,

$$\mathbb{E}\left\{\max_{k=1,\cdots,N} \log \lambda_{\max}(\boldsymbol{H}_k)\right\} \lesssim \log\left(\log N + O(\log\log N)\right) \tag{20}$$

Combining (16), (18), and (20), we get

$$\mathcal{R}_{\text{Opt}} - \mathcal{R}_{\text{Prop}} \;\; \geq \;\; (M-1)\log\frac{\log N}{t} + O\left(\frac{\log\log N}{\log N}\right) + O\left(\frac{1}{t}\right). \tag{21}$$

Consequently, for $\lim_{N\to\infty} t = \infty$ and $\lim_{N\to\infty} \dfrac{t}{\log N} < 1$, $\lim_{N\to\infty} \mathcal{R}_{\text{Opt}} - \mathcal{R}_{\text{Prop}} \neq 0$.

***Case II;*** $\lim_{N\to\infty} t = c$, *where* $c$ *is a constant*: In this case, (16) can be written as

$$\begin{aligned}\mathcal{R}_{\text{Opt}} - \mathcal{R}_{\text{Prop}} \;\; &\geq \;\; M\log\left(\frac{P}{M}\log N + O(\log\log N)\right) - \sum_{i=1}^{M}\mathbb{E}\left\{\log(1+P\lambda_{d_i}(s_i))\right\} \\ &\geq \;\; M\log\left(\frac{P}{M}\log N + O(\log\log N)\right) - \mathbb{E}\left\{\log\left(1 + P\max_{k=1,\cdots,N}\lambda_{\max}(\boldsymbol{H}_k)\right)\right\} \\ &\quad\; -(M-1)\mathbb{E}\left\{\log(1+P\lambda)/\lambda > t\right\}.\end{aligned} \tag{22}$$

Similar to (20), it is easy to see that

$$\mathbb{E}\left\{\log\left(1 + P\max_{k=1,\cdots,N}\lambda_{\max}(\boldsymbol{H}_k)\right)\right\} \lesssim \log P + \log(\log N + O(\log\log N)). \tag{23}$$

Indeed, since $\mathbb{E}\left\{\log(1+P\lambda)\right\} < \infty$, we have $\mathbb{E}\left\{\log(1+P\lambda)/\lambda > t\right\} \sim O(1)$. Hence,

$$\mathcal{R}_{\text{Opt}} - \mathcal{R}_{\text{Prop}} \;\; \geq \;\; (M-1)\log\log N + O(1). \tag{24}$$

As a result, $\lim_{N\to\infty} \mathcal{R}_{\text{Opt}} - \mathcal{R}_{\text{Prop}} \neq 0$. This completes the proof of

$$\lim_{N\to\infty}\frac{t}{\log N} < 1 \Rightarrow \lim_{N\to\infty}\mathcal{R}_{\text{Opt}} - \mathcal{R}_{\text{Prop}} \neq 0.$$

***Case III;*** $\lim_{N\to\infty} \dfrac{t}{\log N} > 1$**:** Let us define $p_k$ as the probability that the maximum singular value of a randomly chosen user $k$ is greater than $t$. In [16], it is shown that for a $K \times M$ matrix $\boldsymbol{A}$, whose entries are i.i.d Gaussian with zero mean and variance one, we have

$$\text{Prob}\{\lambda_{\max}(\boldsymbol{A}) > t\} \sim \frac{t^{M+K-2}\exp(-t)}{\Gamma(M)\Gamma(K)}\left[1 + O\left(t^{-1}\right)\right]. \tag{25}$$

Therefore,

$$p_k \;\; = \;\; \frac{t^{M+K-2}\exp(-t)}{\Gamma(M)\Gamma(K)}\left[1 + O\left(t^{-1}\right)\right], \tag{26}$$

which is independent of $k$, and we denote it with $p$. We define $L$ as the number of users whose maximum singular values are greater than $t$. Since $L$ is a binomial random variable with parameter $p$, $\mathbb{E}\{L\} = Np$.

Using [28], Theorem 1, we can write

$$\mathcal{R}_{\rm Opt} - \mathcal{R}_{\rm Prop} \geq (1-p)^N (\mathcal{R}_1 - \mathcal{R}_{\mathcal{A}}^{\rm NCSI}), \tag{27}$$

where $\mathcal{R}_1 = \mathbb{E}\left\{ \max_{\substack{\boldsymbol{Q}_n \\ \sum {\rm Tr}(\boldsymbol{Q}_{\rm n})={\rm P}}} \log\det\left(\boldsymbol{I}_M + \sum_{n=1}^N \boldsymbol{H}_n^* \boldsymbol{Q}_n \boldsymbol{H}_n\right) \middle| \mathcal{A}\right\}$, $\mathcal{A}$ is the event that $L = 0$, and $\mathcal{R}_{\mathcal{A}}^{\rm NCSI}$ stands for the sum-rate of MIMO-BC when no CSI is available at the base station, conditioned on $\mathcal{A}$. In [29], it has been shown that

$$\mathcal{R}^{\rm NCSI} = \mathbb{E}_{\boldsymbol{H}_k}\left\{\log\det\left[\boldsymbol{I} + \frac{P}{M}\boldsymbol{H}_k\boldsymbol{H}_k^*\right]\right\}. \tag{28}$$

Since $\lim_{N\to\infty} \frac{t}{\log N} > 1$, using (26), it can be easily shown that $Np \to 0$. As a result, with a similar approach as in [28], we have

$$\begin{aligned} \mathcal{R}_{\mathcal{A}}^{\rm NCSI} &= \mathbb{E}_{\boldsymbol{H}_k|\mathcal{A}}\left\{\log\det\left[\boldsymbol{I} + \frac{P}{M}\boldsymbol{H}_k\boldsymbol{H}_k^*\right] \middle| \mathcal{A}\right\} \\ &\sim O(1). \end{aligned} \tag{29}$$

Indeed, we can write

$$\begin{aligned} \mathcal{R}_1 &\geq \mathbb{E}\left\{\log(1+P\theta_{\max})\,|\theta_{\max} < t\right\} \\ &\geq \mathbb{E}\left\{\log(1+P\theta_{\max})\,|\theta_{\max} < t, \theta_{\max} > \log N\right\} \text{Prob}\{\theta_{\max} > \log N|\theta_{\max} < t\} \\ &\geq \log(1+P\log N)\vartheta, \end{aligned} \tag{30}$$

where $\theta_{\max} \triangleq \max_k \lambda_{\max}(\boldsymbol{H}_k)$, and $\vartheta \triangleq$ Prob$\{\theta_{\max} > \log N|\theta_{\max} < t\}$. Using (26), $\vartheta$ can be written as follows:

$$\vartheta = \frac{\left(1 - \frac{t^{M+K-2}e^{-t}(1+O(t^{-1}))}{\Gamma(M)\Gamma(K)}\right)^N - \left(1 - \frac{[\log N]^{M+K-2}(1+O([\log N]^{-1}))}{N\Gamma(M)\Gamma(K)}\right)^N}{\left(1 - \frac{t^{M+K-2}e^{-t}(1+O(t^{-1}))}{\Gamma(M)\Gamma(K)}\right)^N}. \tag{31}$$

Since $\lim_{N\to\infty}\frac{t}{\log N}>1$, it can be shown that $\vartheta\sim 1-o(\frac{1}{N})$. Substituting $\vartheta$ in (30), yields

$$\mathcal{R}_1 \;\geq\; \log(1+P\log N)\left(1-o\Big(\frac{1}{N}\Big)\right). \tag{32}$$

Using the above equation and (29), the right hand side of (27) can be lower-bounded as,

$$\begin{aligned}\text{RH}(27) &\geq (1-p)^N[\log\log N+O(1)]\\ &\sim e^{-Np(1+O(p))}[\log\log N+O(1)]\\ &\sim \log\log N.\end{aligned} \tag{33}$$

The last line in the above equation follows from $\lim_{N\to\infty}\frac{t}{\log N}>1$, which incurs $Np\to 0$. As a result, $\mathcal{R}_{\text{Opt}}-\mathcal{R}_{\text{Prop}}\neq 0$. This completes the proof for the necessity of $\rho(N)\sim o(\log N)$.

*The necessity of* $\rho(N)=\log\log\log\log N+\log[\Gamma(K)\Gamma(M)]+\omega\left(\frac{1}{\log\log\log N}\right)$:

Let $\rho(N)=\log\log\log\log N+\log[\Gamma(M)\Gamma(K)]+\sigma(N)$. Suppose that

$$\rho(N)\nsim \log\log\log\log N+\log[\Gamma(K)\Gamma(M)]+\omega\left(\frac{1}{\log\log\log N}\right), \tag{34}$$

which incurs $\sigma(N)\sim O\left(\frac{1}{\log\log\log N}\right)$, or $\sigma(N)<0$. Using (27), we have

$$\mathcal{R}_{\text{Opt}}-\mathcal{R}_{\text{Prop}} \;\geq\; (1-p)^N[\mathcal{R}_1-\mathcal{R}_{\mathcal{A}}^{\text{NCSI}}]. \tag{35}$$

Similar to (29) and (32), under the assumption of (34), it can be shown that

$$\begin{aligned}\mathcal{R}_1 &\geq \log(1+P\log N)\left(1-o\Big(\frac{1}{N}\Big)\right),\\ \mathcal{R}_{\mathcal{A}}^{\text{NCSI}} &\sim O(1).\end{aligned} \tag{36}$$

Using the above equations and (26), we can write

$$\begin{aligned}
\mathcal{R}_{\text{Opt}} - \mathcal{R}_{\text{Prop}} &\geq \left(1 - \frac{t^{M+K-2}\exp(-t)}{\Gamma(M)\Gamma(K)}\left[1 + O(t^{-1})\right]\right)^N [\log\log N + O(1)] \\
&\sim \left(1 - \frac{e^{\rho(N)}}{N\Gamma(M)\Gamma(K)}\left[1 + O\left(\frac{\log\log N}{\log N}\right)\right]\right)^N [\log\log N + O(1)] \\
&\sim \exp\left\{-\frac{e^{\rho(N)}}{\Gamma(M)\Gamma(K)}\right\}[1 + o(1)]\,[\log\log N + O(1)] \\
&\sim \exp\left\{-e^{\sigma(N)}\log\log\log N\right\}[\log\log N + O(1)]\,[1 + o(1)] \\
&\sim M\exp\left\{[1 - e^{\sigma(N)}]\log\log\log N\right\}[1 + o(1)]\,. \qquad (37)
\end{aligned}$$

Under the assumption of (34), in the case of $\sigma(N) = O\left(\frac{1}{\log\log\log N}\right)$, i.e.,

$$\lim_{N\to\infty} \sigma(N)\log\log\log N = c < \infty,$$

using (37), we have

$$\begin{aligned}
\mathcal{R}_{\text{Opt}} - \mathcal{R}_{\text{Prop}} &\geq \exp\left\{[-\sigma(N) + O(\sigma^2(N))]\log\log\log N\right\}[1 + o(1)] \\
&\sim \exp\{-\sigma(N)\log\log\log N\}[1 + o(1)]\,. \qquad (38)
\end{aligned}$$

Hence,

$$\begin{aligned}
\lim_{N\to\infty} \mathcal{R}_{\text{Opt}} - \mathcal{R}_{\text{Prop}} &\geq e^{-c} \\
&\neq 0. \qquad (39)
\end{aligned}$$

Also, in the case of $\sigma(N) < 0$, using (37), we have

$$\begin{aligned}
\lim_{N\to\infty} \mathcal{R}_{\text{Opt}} - \mathcal{R}_{\text{Prop}} &\geq 1 \\
&\neq 0. \qquad (40)
\end{aligned}$$

This completes the proof for the necessity of $\rho(N) = \log\log\log\log N + \log[\Gamma(K)\Gamma(M)] + \omega\left(\frac{1}{\log\log\log N}\right)$.

■

**Theorem 2** *The sufficient condition to achieve* $\lim_{N\to\infty}\mathcal{R}_{\text{Opt}}-\mathcal{R}_{\text{Prop}}=0$ *is having*

$$t=\log N+(M+K-2)\log\log N-\rho(N), \tag{41}$$

*where* $\rho(N)$ *satisfies*

$$\rho(N) \quad \sim \quad o(\log N),$$

*and*

$$\rho(N) \quad \sim \quad \log\log\log\log N+\omega\left(1\right).$$

**Proof -** First, we state and prove the following lemmas:

***Lemma 1****- Assuming* $K>1$*, define* $\Omega_J$ *as the probability of existing at least one user from which* $J$ *eigenvectors (*$J>1$*) are selected in Algorithm 1. Setting* $t=\log N+(M+K-2)\log\log N-\rho(N)$*, in which* $\rho(N)$ *satisfies the conditions of Theorem 2, we have*

$$\Omega_J\sim O\left(\frac{e^{o(\log N)}}{N^{J-1}}\right). \tag{42}$$

**Proof**- Consider the following event [3]:

$$A_k=\{\lambda_i(k)>t, \quad i=1,\cdots J, \quad \lambda_i(k)<t, \quad i=J+1,\cdots,K\}. \tag{43}$$

[3]We have assumed that the singular values are in the decreasing order, i.e., $\lambda_1>\lambda_2>\cdots>\lambda_K$

We have

$$\begin{aligned} \|\boldsymbol{H}_k\|^2 &= \mathrm{Tr}\{\boldsymbol{H}_\mathrm{k}\boldsymbol{H}_\mathrm{k}^*\} \\ &= \sum_{i=1}^{K}\lambda_i(k) \\ &\geq \sum_{i=1}^{J}\lambda_i(k). \end{aligned} \tag{44}$$

Since $t = \log N + o(\log N)$, we can write

$$\text{Prob}\{A_k\} \;\leq\; \text{Prob}\{\|\boldsymbol{H}_k\|^2 \geq J\log N + o(\log N)\}, \tag{45}$$

As $\|\boldsymbol{H}_k\|^2$ has a chi-square distribution with $2MK$ degrees of freedom [30], the right hand side of (45) can be written as

$$\begin{aligned} \text{Prob}\left\{\|\boldsymbol{H}_k\|^2 \geq J\log N + o(\log N)\right\} &= \int_{J\log N + o(\log N)}^{\infty} \frac{x^{MK-1}\exp(-x)}{\Gamma(MK)}dx \\ &= \sum_{m=0}^{MK-1} \frac{[J\log N + o(\log N)]^m}{m!} e^{-J\log N + o(\log N)} \\ &= \frac{\left([J\log N]^{MK-1} + o([\log N]^{MK-1})\right) e^{o(\log N)}}{N^J (MK-1)!} \\ &= \Psi_J \frac{[\log N]^{MK-1} e^{o(\log N)}}{N^J}[1+o(1)], \end{aligned} \tag{46}$$

where $\Psi_J = \frac{J^{MK-1}}{(MK-1)!}$. Using (45), and (46), we can write $\Omega_J$ as

$$\begin{aligned} \Omega_J &= 1 - \prod_{k=1}^{N}(1 - \text{Prob}\{A_k\}) \\ &\leq 1 - \left[1 - \Psi_J \frac{[\log N]^{MK-1} e^{o(\log N)}}{N^J}[1+o(1)]\right]^N \\ &\sim 1 - \exp\left\{N\log\left[1 - \Psi_J \frac{[\log N]^{MK-1} e^{o(\log N)}}{N^J}[1+o(1)]\right]\right\} \\ &\sim 1 - \exp\left\{-\Psi_J \frac{[\log N]^{MK-1} e^{o(\log N)}}{N^{J-1}}[1+o(1)]\right\} \\ &\sim O\left(\frac{e^{o(\log N)}}{N^{J-1}}\right). \end{aligned} \tag{47}$$

■

As a result, $\lim_{N\to\infty}\Omega_J = 0$, for $J > 1$. This implies that as $N \to \infty$, with probability one, at most one eigenvector for each user is likely to be selected by this algorithm. This eigenvector corresponds to the maximum singular value of that user.

***Lemma 2-*** *Let* $t = \log N + (M+K-2)\log\log N - \rho(N)$*, in which* $\rho(N)$ *satisfies the conditions of Theorem 2, and* $L$ *be the number of users being selected in the first step of Algorithm 1. Then, as* $N \to \infty$*, with probability one*

$$L \sim \frac{e^{\rho(N)}}{\Gamma(M)\Gamma(K)}\left[1 + o\left(1\right)\right]. \tag{48}$$

**Proof-** Using (26), the probability of a randomly chosen user $k$ being pre-selected in the first step of Algorithm 1 can be calculated as,

$$\begin{aligned} p &= \text{Prob}\{\lambda_{\max}(\boldsymbol{H}_k) > t\} \\ &\sim \frac{t^{M+K-2}e^{-t}}{\Gamma(M)\Gamma(K)}\left(1 + O(t^{-1})\right) \\ &\sim \frac{e^{\rho(N)}}{N\Gamma(M)\Gamma(K)}\left[1 + o(1)\right] \\ &\sim \frac{\log\log\log N e^{q(N)}}{N\Gamma(M)\Gamma(K)}\left[1 + o(1)\right], \end{aligned} \tag{49}$$

where $q(N) = \rho(N) - \log\log\log\log N$. Consider the following probability:

$$\xi = \text{Prob}\left\{Np(1-\epsilon) < L < Np(1+\epsilon)\right\}, \tag{50}$$

where $\epsilon=\sqrt{2\Gamma(M)\Gamma(K)e^{\frac{-q(N)}{4}}}$. Note that since $q(N)=\omega(1)$, we have $\lim_{N\to\infty}\epsilon=0$. $\xi$ can be computed as

$$\begin{aligned}\xi &= \sum_{l=\lceil Np(1-\epsilon)\rceil}^{\lfloor Np(1+\epsilon)\rfloor}\binom{N}{l}p^l(1-p)^{N-l}\\ &\approx 1-Q\left(\frac{Np-Np(1-\epsilon)}{\sqrt{Np(1-p)}}\right)-Q\left(\frac{Np(1+\epsilon)-Np}{\sqrt{Np(1-p)}}\right)\\ &= 1-2Q\left(\frac{\sqrt{Np}\epsilon}{\sqrt{1-p}}\right)\\ &\approx 1-\frac{2\sqrt{1-p}}{\sqrt{2\pi}\sqrt{Np}\epsilon}\exp\left(-\frac{Np\epsilon^2}{2(1-p)}\right). \end{aligned} \tag{51}$$

Substituting $p$ from (49), and having $\epsilon^2=2\Gamma(M)\Gamma(K)e^{\frac{-q(N)}{2}}$, we have

$$\xi \sim 1-O\left(\frac{e^{\frac{-q(N)}{4}}}{\sqrt{\log\log\log N}}\right)\exp\left\{-\log\log\log Ne^{\frac{q(N)}{2}}[1+o(1)]\right\} \tag{52}$$

Thus, $\lim_{N\to\infty}\xi=1$. Finally, using (49) and (52), with probability one we have

$$\begin{aligned} L &\sim Np\left(1+O(\epsilon)\right)\\ &\sim \frac{e^{\rho(N)}}{\Gamma(M)\Gamma(K)}\left[1+o\left(1\right)\right]. \end{aligned} \tag{53}$$

■

Since $\rho(N)=o(\log N)$, from Lemma 2, it is evident that $\lim_{N\to\infty}\frac{L}{N}=0$. Therefore, only a small fraction of users are pre-selected. This results in reducing the amount of feedback sent to the base station.

As shown in Lemma 1, in the asymptotic case of $N\to\infty$, at most one eigenvector from each user is likely to be selected. This eigenvector corresponds to the maximum singular value of that user's channel matrix, and is denoted by $\boldsymbol{V}_{i,\max}$. Hence, for the sake of simplicity of notation, we define the measure of orthogonality between the users $i$ and $j$, denoted by $\mathcal{O}(i,j)$, as the orthogonality measure between $\boldsymbol{V}_{i,\max}$ and $\boldsymbol{V}_{j,\max}$, defined in (3) as $z\left(\boldsymbol{V}_{i,\max},\boldsymbol{V}_{j,\max}\right)$.

In other words,

$$\mathcal{O}(i,j) = |\boldsymbol{V}_{i,\max}^{*}\boldsymbol{V}_{j,\max}|^2. \tag{54}$$

***Lemma 3-*** *The probability density function of $\mathcal{O}(i,j)$ defined in (54) can be computed from*

$$p_{\mathcal{O}(i,j)}(z) = (M-1)(1-z)^{M-2}. \tag{55}$$

***Proof-*** In Appendix A.

***Definition1-*** *A set $\mathcal{S} = \{\boldsymbol{h}_i\}_{i=1}^{M}$, in which $\boldsymbol{h}_i \in \mathbb{C}^{1\times M}$, is called $\epsilon$-orthogonal if we have $z(\boldsymbol{h}_i, \boldsymbol{h}_j) < \epsilon$, for every $\boldsymbol{h}_i \neq \boldsymbol{h}_j \in \mathcal{S}$.*

***Lemma 4-*** *Let $t = \log N + (M+K-2)\log\log N - \rho(N)$, where $\rho(N)$ satisfies the conditions of Theorem 2. Then, as $N \to \infty$, the selected coordinates by Algorithm 1 construct an $\epsilon(N)$-orthogonal set, with probability one, where $\epsilon(N) = e^{-\frac{q(N)}{M}}$, and $q(N) = \rho(N) - \log\log\log\log N$.*

**Proof-** After selecting the first user, $s_1$, with largest maximum singular value, the user which is most orthogonal to $s_1$ is selected. In other words,

$$s_2 = \arg \min_{l\in\mathcal{S}_1} \mathcal{O}(l, s_1), \tag{56}$$

where $\mathcal{S}_1$ is defined in (7). First, we show that the users $s_1$ and $s_2$ are with probability one $\epsilon(N)$-orthogonal to each other, or equivalently, $\mathcal{O}(s_2, s_1) < \epsilon(N)$. To do this, consider the following probability:

$$\mu = \text{Prob}\left\{\mathcal{O}(s_2, s_1) < \epsilon(N)\right\}. \tag{57}$$

Using (55), this probability can be written as

$$\begin{aligned}\mu &= \text{Prob}\left\{\min_l \mathcal{O}(l, s_1) < \epsilon(N)\right\} \\ &= 1 - \left(\text{Prob}\left\{\mathcal{O}(l, s_1) > \epsilon(N)\right\}\right)^{L-1} \\ &= 1 - \left(\int_{\epsilon(N)}^{1} (M-1)(1-z)^{M-2} dz\right)^{L-1} \\ &= 1 - [1-\epsilon(N)]^{(L-1)(M-1)} \\ &= 1 - \exp\left\{-(L-1)(M-1)\log\left[1-\epsilon(N)\right]\right\} \\ &= 1 - \exp\left\{-(L-1)(M-1)\left[\epsilon(N) + O\left(\epsilon^2(N)\right)\right]\right\}. \end{aligned} \quad (58)$$

Defining the event $\mathcal{D} = \{Np(1-\epsilon) < L < Np(1+\epsilon)\}$, with $p$ and $\epsilon$ defined in (49) and (50), and using (52), a lower bound for $\mu$ is found as,

$$\begin{aligned}\mu &\geq \text{Prob}\{\mathcal{D}\}\left[1 - \exp\left\{-(Np(1-\epsilon)-1)(M-1)\left[\epsilon(N) + O\left(\epsilon^2(N)\right)\right]\right\}\right] \\ &\sim \left[1 - O\left(\frac{e^{\frac{-q(N)}{4}}}{\sqrt{\log\log\log N}}\right)\exp\left\{-\log\log\log N e^{\frac{q(N)}{2}}[1+o(1)]\right\}\right] \times \\ &\quad \left[1 - \exp\left\{-\frac{\log\log\log N e^{\frac{(M-1)q(N)}{M}}}{\Gamma(M-1)\Gamma(K)}[1+o(1)]\right\}\right]. \end{aligned} \quad (59)$$

Since $q(N) \sim \omega(1)$, the above probability approaches one as $N \to \infty$. Therefore, with probability one users $s_1$ and $s_2$ are $\epsilon(N)$-orthogonal to each other.

Now, assume that $m$ users, which construct an $\epsilon(N)$-orthogonal set $\mathcal{A}_m$, are selected up to the $m$th step of Algorithm 1. We show that the selected user in the $(m+1)$th step of this algorithm, $s_{m+1}$, is such that with probability one, $\mathcal{A}_{m+1} = \mathcal{A}_m + \{s_{m+1}\}$ is $\epsilon(N)$-orthogonal, or equivalently, $s_{m+1}$ is $\epsilon(N)$-orthogonal to all users in $\mathcal{A}_m$. To this end, we define the following probability:

$$\nu_{k,m} = \text{Prob}\{\mathcal{O}(s_1, k) < \alpha, \mathcal{O}(s_2, k) < \alpha, \cdots, \mathcal{O}(s_m, k) < \alpha\}, \quad (60)$$

where $\alpha = \frac{\epsilon(N)}{M}$. $\nu_{k,m}$ is the probability that a randomly selected user $k$ is $\alpha$-orthogonal to all users in $\mathcal{A}_m$. This probability can be written as

$$\nu_{k,m} = \text{Prob}\left\{\mathcal{O}(s_1, k) < \alpha\right\} \prod_{i=2}^{m} \kappa_i, \tag{61}$$

where $\kappa_i = \text{Prob}\left\{\mathcal{O}(s_i, k) < \alpha \mid \quad \mathcal{O}(s_1, k) < \alpha, \cdots, \mathcal{O}(s_{i-1}, k) < \alpha\right\}$. From (55), the first term in the right hand side of the above equation can be written as

$$\begin{aligned} \text{Prob}\left\{\mathcal{O}(s_1, k) < \alpha\right\} &= \int_0^{\alpha} (M-1)(1-z)^{M-2} dz \\ &= 1 - (1-\alpha)^{M-1} \\ &\sim (M-1)\alpha + O(\alpha^2). \end{aligned} \tag{62}$$

In Appendix B, it has been proved that

$$\kappa_i \sim (M-i)\alpha + O(\alpha^{3/2}). \tag{63}$$

Hence, using (61), (62), and (63), we can write

$$\begin{aligned} \nu_{k,m} &\sim \left[(M-1)\alpha + O(\alpha^2)\right] \prod_{i=2}^{m} \left[(M-i)\alpha + O(\alpha^{3/2})\right] \\ &\sim \frac{\Gamma(M)}{\Gamma(M-m)} \alpha^m + O\left(\alpha^{m+1/2}\right) \\ &\sim \frac{\Gamma(M)}{\Gamma(M-m)M^m} \left[[\epsilon(N)]^m + O\left([\epsilon(N)]^{(m+1/2)}\right)\right] \end{aligned} \tag{64}$$

Now, we define $\omega_m$ as the probability of existing at least one user $\alpha$-orthogonal to the users in the set $\mathcal{A}_m$. Noting that $\nu_{k,m}$ is the same for all $k$, we obtain,

$$\begin{aligned} \omega_m &= 1 - \prod_{k=1}^{L-m} (1 - \nu_{k,m}) \\ &= 1 - \exp\left\{(L-m)\log(1-\nu_{k,m})\right\} \\ &= 1 - \exp\left\{(L-m)\left[-\nu_{k,m} + O(\nu_{k,m}^2)\right]\right\}. \end{aligned} \tag{65}$$

Similar to (59), we can compute $\omega_m$ as,

$$\omega_m \sim \left[1 - O\left(\frac{e^{\frac{-q(N)}{4}}}{\sqrt{\log\log\log N}}\right)\exp\left\{-\log\log\log N e^{\frac{q(N)}{2}}[1+o(1)]\right\}\right] \times \left[1 - \exp\left\{-\frac{\log\log\log N e^{\frac{(M-m)q(N)}{M}}}{\Gamma(M-m)M^m\Gamma(K)}[1+o(1)]\right\}\right]. \quad (66)$$

Since $m \leq M-1$, it follows that $\lim_{N\to\infty}\omega_m = 1$. In other words, as $N$ tends to infinity, with probability one there exists at least one user $u_{m+1}$, $\alpha$-orthogonal to all users in $\mathcal{A}_m$.

Consider user $s_{m+1}$ which is selected in the $(m+1)$th step of Algorithm 1. Obviously, we have

$$\begin{aligned}\sum_{j=1}^{m}\mathcal{O}(s_{m+1}, s_j) &\leq \sum_{j=1}^{m}\mathcal{O}(u_{m+1}, s_j) \\ &\leq m\alpha \\ &= \frac{m\epsilon(N)}{M} \\ &\leq \epsilon(N). \end{aligned} \quad (67)$$

Knowing the fact that $\mathcal{O}(s_{m+1}, s_j) \geq 0$, for $j = 1, \cdots m$, we can write

$$\mathcal{O}(s_{m+1}, s_j) \leq \epsilon(N), \quad j = 1, \cdots m$$

which means that with probability one, $s_{m+1}$ is $\epsilon(N)$-orthogonal to the users in the set $\mathcal{A}_m$, and consequently, $\mathcal{A}_{m+1}$ is an $\epsilon(N)$-orthogonal set.

Let us define $\mathcal{X}_m$ as the event that the set $\mathcal{A}_m$ is $\epsilon(N)$-orthogonal. We can write

$$\text{Prob}\{\mathcal{X}_M\} = \text{Prob}\{\mathcal{X}_2\}\prod_{m=3}^{M}\text{Prob}\{\mathcal{X}_m|\mathcal{X}_{m-1}\}. \quad (68)$$

From (59) and (66), the above probability is lower-bounded as

$$\begin{aligned}
\text{Prob}\{\mathcal{X}_M\} &\geq \mu \prod_{m=2}^{M-1} \omega_m \\
&\geq \left[1 - O\left(\frac{e^{\frac{-q(N)}{4}}}{\sqrt{\log\log\log N}}\right) \exp\left\{-\log\log\log N e^{\frac{q(N)}{2}} [1+o(1)]\right\}\right]^{M-1} \times \\
&\quad \left[1 - \exp\left\{-\frac{\log\log\log N e^{\frac{(M-1)q(N)}{M}}}{\Gamma(M-1)\Gamma(K)} [1+o(1)]\right\}\right] \times \\
&\quad \prod_{m=2}^{M-1} \left[1 - \exp\left\{-\frac{\log\log\log N e^{\frac{(M-m)q(N)}{M}}}{\Gamma(M-m)M^m\Gamma(K)} [1+o(1)]\right\}\right] \\
&\sim 1 - \exp\left\{-\frac{\log\log\log N e^{\frac{q(N)}{M}}}{\Gamma(M-m)M^m\Gamma(K)} [1+o(1)]\right\}. \qquad (69)
\end{aligned}$$

Therefore, $\lim_{N\to\infty} \text{Prob}\{\mathcal{X}_M\} = 1$. In other words, the selected coordinates by Algorithm 1, with probability one, construct an $\epsilon(N)$-orthogonal set as $N$ tends to infinity, which completes the proof of Lemma 4.

■

As mentioned earlier, after selecting the coordinates, the "selected coordinate matrix", $\boldsymbol{\mathcal{H}}$, is constructed using (9). By applying zero-forcing beam-forming, the information vector, $\boldsymbol{u}$, is multiplied by $\boldsymbol{\mathcal{H}}^{-1}$ to construct the transmitted signal as (10). Using (11), we can write

$$\boldsymbol{r} = \boldsymbol{u} + \boldsymbol{w}, \qquad (70)$$

where $\boldsymbol{r} = [r_{s_1,d_1}, \cdots, r_{s_M,d_M}]^T$, $\boldsymbol{u} = [u_{s_1,d_1}, \cdots, u_{s_M,d_M}]^T$, and $\boldsymbol{w} = [w_{s_1,d_1}, \cdots, w_{s_M,d_M}]^T$. Having the power constraint $P$ for $\boldsymbol{x}$, the sum-rate capacity can be computed as [19],

$$\mathcal{R}_{\text{Prop}} = \mathbb{E}_{\boldsymbol{\mathcal{H}}}\left\{ \max_{\substack{P_m \\ \sum_{m=1}^{M} \gamma_m P_m \leq P}} \sum_{m=1}^{M} \log(1+P_m) \right\}, \quad \gamma_m = \left[(\boldsymbol{\mathcal{H}}^*\boldsymbol{\mathcal{H}})^{-1}\right]_{m,m}, \qquad (71)$$

where $[\boldsymbol{A}]_{i,j}$ denotes the entry of matrix $\boldsymbol{A}$ in the $i$th row and the $j$th column. The optimal $P_m$'s in (71) can be obtained by "water-filling". Here, we assume that $P_m$'s are all equal (uniform power allocation). Thus,

$$P_m = \frac{P}{\mathrm{Tr}\left\{[\boldsymbol{\mathcal{H}}^*\boldsymbol{\mathcal{H}}]^{-1}\right\}}. \tag{72}$$

Consequently,

$$\mathcal{R}_{\text{Prop}}^{\text{U}} = \mathbb{E}_{\boldsymbol{\mathcal{H}}}\left\{M\log\left(1+\frac{P}{\mathrm{Tr}\left\{[\boldsymbol{\mathcal{H}}^*\boldsymbol{\mathcal{H}}]^{-1}\right\}}\right)\right\}, \tag{73}$$

where $\mathcal{R}_{\text{Prop}}^{\text{U}}$ stands for the sum-rate achieving by the proposed method, when the power is uniformly allocated among the coordinates.

Having defined $\mathcal{X}_M$ in (68) and using (69), the above equation can be written as follows:

$$\begin{aligned}
\mathcal{R}_{\text{Prop}}^{\text{U}} &= \mathbb{E}_{\boldsymbol{\mathcal{H}}}\left\{M\log\left(1+\frac{P}{\mathrm{Tr}\left\{[\boldsymbol{\mathcal{H}}^*\boldsymbol{\mathcal{H}}]^{-1}\right\}}\right)\middle|\,\mathcal{X}_M\right\}\text{Prob}\{\mathcal{X}_M\}+\\
&\quad \mathbb{E}_{\boldsymbol{\mathcal{H}}}\left\{M\log\left(1+\frac{P}{\mathrm{Tr}\left\{[\boldsymbol{\mathcal{H}}^*\boldsymbol{\mathcal{H}}]^{-1}\right\}}\right)\middle|\,\mathcal{X}_M^C\right\}(1-\text{Prob}\{\mathcal{X}_M\})\\
&\geq \mathbb{E}_{\boldsymbol{\mathcal{H}}}\left\{M\log\left(1+\frac{P}{\mathrm{Tr}\left\{[\boldsymbol{\mathcal{H}}^*\boldsymbol{\mathcal{H}}]^{-1}\right\}}\right)\middle|\,\mathcal{X}_M\right\}\text{Prob}\{\mathcal{X}_M\}\\
&\sim \left(1-\exp\left\{-\frac{\log\log\log N e^{\frac{q(N)}{M}}}{\Gamma(M-m)M^m\Gamma(K)}[1+o(1)]\right\}\right)\times\\
&\quad \mathbb{E}_{\boldsymbol{\mathcal{H}}}\left\{M\log\left(1+\frac{P}{\mathrm{Tr}\left\{[\boldsymbol{\mathcal{H}}^*\boldsymbol{\mathcal{H}}]^{-1}\right\}}\right)\middle|\,\mathcal{X}_M\right\},
\end{aligned} \tag{74}$$

where $\mathcal{X}_M^C$ is the complement of $\mathcal{X}_M$.

From Algorithm 1, it is obvious that the corresponding singular values of the selected eigenvectors are greater that $t = \log N + (M+K-2)\log\log N - \rho(N)$. However, the following lemma which is proved in Appendix C, states that the singular values of all selected dimensions, with probability one, can not exceed $\log N + (M+K-1)\log\log N$:

***Lemma 5***- *Let* $t = \log N + (M+K-1)\log\log N$. *Then,*

$$\eta = \text{Prob}\left\{\max_{k=1,\cdots,N}\lambda_{\max}(\boldsymbol{H}_k) > t\right\} = O\left(\frac{1}{\log N}\right). \tag{75}$$

As a result of this lemma, the singular values corresponding to the all selected dimensions can be expressed as $\log N + o(\log N)$.

To compute the conditional probability $\mathbb{E}_{\boldsymbol{\mathcal{H}}}\left\{ M \log\left(1 + \frac{P}{\mathrm{Tr}\left\{[\boldsymbol{\mathcal{H}}^*\boldsymbol{\mathcal{H}}]^{-1}\right\}}\right) \Big| \mathcal{X}_M \right\}$, we define $\boldsymbol{\mathcal{B}} = \boldsymbol{\mathcal{H}}\boldsymbol{\mathcal{H}}^*$. Conditioned on $\mathcal{X}_M$, i.e., having $\epsilon(N)$-orthogonality among the selected dimensions, using (9), and the results of Lemma 4 and Lemma 5, we can write

$$\boldsymbol{\mathcal{B}}_{ii} = \|\boldsymbol{g}_{s_i,d_i}\|^2 \sim \log N + f(N), \tag{76}$$

and

$$\begin{aligned} |\boldsymbol{\mathcal{B}}_{ij}| &= \sqrt{\|\boldsymbol{g}_{s_i,d_i}\|^2 \|\boldsymbol{g}_{s_j,d_j}\|^2 z\left(\boldsymbol{V}_{s_i,d_i}, \boldsymbol{V}_{s_j,d_j}\right)} \\ &\sim \sqrt{O(\log N) \times O(\log N) \times O\left(\epsilon(N)\right)} \\ &\sim O(\epsilon(N)\log N), \end{aligned} \tag{77}$$

where $f(N) \sim o(\log N)$. In Appendix D it has been shown that any diagonal element of $\boldsymbol{\mathcal{B}}^{-1}$ can be expressed as $[\log N]^{-1} + O\left(\frac{h(N)}{\log N}\right)$, where

$$h(N) \triangleq \max\left(\frac{f(N)}{\log N}, \epsilon(N)\right) \sim o(1). \tag{78}$$

Having this, and using the fact that $\mathrm{Tr}\left\{[\boldsymbol{\mathcal{H}}^*\boldsymbol{\mathcal{H}}]^{-1}\right\} = \mathrm{Tr}\left\{\boldsymbol{\mathcal{B}}^{-1}\right\}$, we can write

$$\begin{aligned} \mathbb{E}_{\boldsymbol{\mathcal{H}}}\left\{ M \log\left(1 + \frac{P}{\mathrm{Tr}\left\{[\boldsymbol{\mathcal{H}}^*\boldsymbol{\mathcal{H}}]^{-1}\right\}}\right) \Bigg| \mathcal{X}_M \right\} &= \mathbb{E}_{\boldsymbol{\mathcal{B}}}\left\{ M \log\left(1 + \frac{P}{\mathrm{Tr}\left\{\boldsymbol{\mathcal{B}}^{-1}\right\}}\right) \Bigg| \mathcal{X}_M \right\} \\ &\sim M \log\left(1 + \frac{P}{M[\log N]^{-1} + O\left(\frac{h(N)}{\log N}\right)}\right) \\ &\sim M \log\left(1 + \frac{P}{M[\log N]^{-1}\left[1 + O\left(h(N)\right)\right]}\right) \\ &\sim M \log\left(\frac{P}{M}\log N + O(h(N)\log N)\right). \end{aligned} \tag{79}$$

From (74) and (79), we have

$$\mathcal{R}_{\text{Prop}}^{\text{U}} \geq M \log\left(\frac{P}{M}\log N + O(h(N)\log N)\right)\left(1 - \exp\left\{-\frac{\log\log\log N e^{\frac{q(N)}{M}}}{\Gamma(M-m)M^m\Gamma(K)}[1+o(1)]\right\}\right). \tag{80}$$

Since adaptive power allocation (using "water-filling") results in higher sum-rate than that of uniform power allocation, we have $\mathcal{R}_{\text{Prop}} \geq \mathcal{R}_{\text{Prop}}^{\text{U}}$. Having the fact that [24]

$$\mathcal{R}_{\text{Opt}} \sim M \log\left(\frac{P}{M}\log N + O(\log\log N)\right), \tag{81}$$

and using (80), we have

$$\begin{aligned} \mathcal{R}_{\text{Opt}} - \mathcal{R}_{\text{Prop}} &\leq M\log\left(\frac{P}{M}\log N + g_1(N)\right) - M\log\left(\frac{P}{M}\log N + g_2(N)\right)(1 - g_3(N)) \\ &= M\log\left(1 + \frac{Mg_1(N)}{P\log N}\right) - M\log\left(1 + \frac{Mg_2(N)}{P\log N}\right) + \\ &\quad Mg_3(N)\log\left(\frac{P}{M}\log N + g_2(N)\right), \end{aligned} \tag{82}$$

where $g_1(N) \sim O(\log\log N)$, $g_2(N) \sim O\left(h(N)\log N\right)$, and

$$g_3(N) \sim \exp\left\{-\frac{\log\log\log N e^{\frac{q(N)}{M}}}{\Gamma(M-m)M^m\Gamma(K)}[1+o(1)]\right\} \sim o\left(\frac{1}{\log\log N}\right). \tag{83}$$

From (78) and (83), and Using the approximation $\log(1+x) \approx x$, for $x \ll 1$, and we can write

$$\begin{aligned} \mathcal{R}_{\text{Opt}} - \mathcal{R}_{\text{Prop}} &\sim M\left(\frac{M[g_1(N) - g_2(N)]}{P\log N}\right) + Mg_3(N)\log\left(\frac{P}{M}\log N + g_1(N)\right) \\ &\sim o(1). \end{aligned} \tag{84}$$

Consequently,

$$\lim_{N\to\infty} \mathcal{R}_{\text{Opt}} - \mathcal{R}_{\text{Prop}} = 0, \tag{85}$$

which completes the proof of Theorem 2.

■

Theorem 2 implies that using Algorithm 1, and applying zero-forcing beam-forming at the base station, the same performance as when the optimum user selection algorithm and optimum precoding scheme is utilized, can asymptotically be achieved.

**Remark 1-** Although in the proof of Theorem 2, we showed that $\lim_{N\to\infty} \mathcal{R}_{\text{Opt}} - \mathcal{R}_{\text{Prop}} = 0$, it is interesting to minimize the order of difference.

Rewriting (84), we get

$$\mathcal{R}_{\text{Opt}} - \mathcal{R}_{\text{Prop}} \sim O\left(\varrho(N)\right) + \exp\left\{-\frac{\log\log\log N e^{1/\epsilon(N)}}{\Gamma(M-m)M^m\Gamma(K)}[1+o(1)]\right\} O(\log\log N), \tag{86}$$

where $\varrho(N) = \max\left(h(N), \frac{\log\log N}{\log N}\right)$, and $h(N)$ is defined in (78). Hence, in order to minimize the order of difference, we must have $h(N) = O\left(\frac{\log\log N}{\log N}\right)$, which incurs $\epsilon(N) = O\left(\frac{\log\log N}{\log N}\right)$ and $f(N) = O(\log\log N)$. As a result,

$$\begin{aligned} q(N) &= -M\log\epsilon(N) \\ &= M\log\log N - M\log\log\log N + \psi(N), \end{aligned} \tag{87}$$

where $\psi(N)$ is an arbitrary function with the condition $\lim_{N\to\infty} \psi(N) = c > 0$. Hence, using the definition of $q(N)$ in Lemma 4, we can write

$$t = \log N + (K-2)\log\log N + M\log\log\log N - \log\log\log\log N - \psi(N). \tag{88}$$

Also, to guarantee $f(N) = O(\log\log N)$, we must have

$$t = \log N + O(\log\log N), \tag{89}$$

which means $\psi(N) \sim O(\log\log N)$. Having these conditions on $t$, we can guarantee $\mathcal{R}_{\text{Opt}} - \mathcal{R}_{\text{Prop}} \sim O\left(\frac{\log\log N}{\log N}\right)$.

**Remark 2-** It is important to note that satisfying $\lim_{N\to\infty} \mathcal{R}_{\text{Opt}} - \mathcal{R}_{\text{Prop}} = 0$, is much more challenging than that of $\lim_{N\to\infty} \frac{\mathcal{R}_{\text{Prop}}}{\mathcal{R}_{\text{Opt}}} = 1$. The following lemma, which is proved in Appendix E, clarifies this fact:

***Lemma 6-*** *Suppose that in Algorithm 1, $t = \log N$, and the coordinates are chosen randomly among the pre-selected eigenvectors. Then,*

$$\lim_{N\to\infty} \frac{\mathcal{R}_{\text{Prop}}}{\mathcal{R}_{\text{Opt}}} = 1. \tag{90}$$

The above lemma states that to satisfy $\lim_{N\to\infty} \frac{\mathcal{R}_{\text{Prop}}}{\mathcal{R}_{\text{Opt}}} = 1$, the orthogonality among the coordinates is not a necessary condition.

*B. Comparison with other Downlink Strategies*

In this section, we compare the performance of our proposed scheme with some other downlink strategies in terms of sum-rate capacity. To have a good measure for comparison, we give the following definition:

***Definition 2-*** *For a MIMO-BC in which a base station, and average power constraint $P$ communicating to $N$ users, using strategy $S$, the* ***multiplexing gain*** *is defined as* [4]

$$r_S = \lim_{P\to\infty} \frac{\mathcal{R}_S(P, N)}{\log P}, \tag{91}$$

*and the* ***multiuser diversity gain*** *is defined as*

$$d_S = \lim_{N\to\infty} \frac{\mathcal{R}_S(P, N)}{r_S \log\log N}, \tag{92}$$

[4]More precisely, as in [31], $r$ is the maximum achievable multiplexing gain when diversity gain approaches zero.

*where $\mathcal{R}_S(P,N)$ is the achievable sum-rate.*

***Lemma 7-*** *Using the proposed algorithm, and applying zero-forcing beam-forming, we can achieve $r = M$, and $d = 1$, which are the maximum achievable values in a MIMO-BC.*

**Proof-** Appendix F.

*1) Time Division Multiple Access (TDMA):* In this scheme, the base station only serves one user in each time slot. Hence, to achieve the maximum sum-rate, the user which has the maximum single-user capacity should be served. Because of its simplicity, this strategy is widely used in the downlink of the cellular networks. The achievable sum-rate of this scheme can be written as

$$\mathcal{R}_{\text{TDMA}} = \mathbb{E}\left\{ \max_{k} \max_{\substack{\boldsymbol{Q}_k \\ \text{Tr}\{\boldsymbol{Q}_\text{k}\}=\text{P}}} \log\det\left[\boldsymbol{I}_{K\times K} + \boldsymbol{H}_k \boldsymbol{Q}_k \boldsymbol{H}_k^*\right] \right\}, \tag{93}$$

where $\boldsymbol{Q}_k$ is obtained by "water-filling". Using (91) and (92), and the result of Lemma 1 in [32], the *multiplexing gain* and *multiuser diversity gain* for this scheme can be obtained as,

$$\begin{aligned} r_{\text{TDMA}} &= \lim_{P\to\infty} \frac{\mathcal{R}_{\text{TDMA}}(P,N)}{\log P} \\ &= \lim_{P\to\infty} \frac{\mathbb{E}\left\{\max_k \left(\sum_{i=1}^{\min(M,K)} \log\left(\frac{P\lambda_i(k)}{\min(M,K)}\right)\right)\right\}}{\log P} \\ &= \min(M,K), \end{aligned} \tag{94}$$

and,

$$\begin{aligned} d_{\text{TDMA}} &= \lim_{N\to\infty} \frac{\mathcal{R}_{\text{TDMA}}(P,N)}{\min(M,K)\log\log N} \\ &= 1. \end{aligned} \tag{95}$$

Hence, this scheme achieves the full multiuser diversity gain, while achieving the full multiplexing gain only in the case of $K \geq M$.

Although this method has been shown to be optimal for single-antenna broadcast channel ($M = 1$) [33], for the case of $M > K \geq 1$, as a result of losing the multiplexing gain, this method is no longer optimum [5].

From the proof of the Lemma 1 in [32], it can be observed that the upper and lower bounds for $\mathcal{R}_{\text{TDMA}}$ have the same behavior asymptotically almost surely, when $N \to \infty$. In other words[6],

$$\begin{aligned} K \log\left(1 + \frac{P}{K} \max_k \lambda_{\min}(\boldsymbol{H}'_k {\boldsymbol{H}'_k}^*)\right) &\sim K \log\left(1 + \frac{P}{K^2} \max_k \operatorname{Tr}(\boldsymbol{H}_{\mathrm{k}} \boldsymbol{H}^*_{\mathrm{k}})\right) \\ &\sim K \log(1 + \frac{P}{K^2} \log N), \end{aligned} \quad (96)$$

where $\boldsymbol{H}'_k$ ($K \times K$) is a truncated version of $\boldsymbol{H}_k$ by omitting the $M - K$ columns of $\boldsymbol{H}_k$. From (96), and having the fact that $\lambda_{\min}(\boldsymbol{H}'_k {\boldsymbol{H}'_k}^*) \leq \lambda_{\min}(\boldsymbol{H}_k {\boldsymbol{H}_k}^*)$, the following observations can be obtained:

**Observation 1-** For the user which maximizes the single-user capacity in (93), ($l$), all the eigenvalues should be of the same order. In other words,

$$\lambda j(\boldsymbol{H}_l \boldsymbol{H}^*_l) \sim \frac{\log N}{K} + O(\log \log N), \qquad j = 1, \cdots, K. \quad (97)$$

As a result of this, $\boldsymbol{H}_l \boldsymbol{H}^*_l$ tends to the identity matrix.

**Observation 2-** The user with maximum single-user capacity has the maximum $\lambda_{\min}$, asymptotically.

For the case of $K \geq M$, similar to (96), the asymptotic sum-rate capacity can be computed

[5]For the case of $K \geq M$, this scheme is not optimal either. This fact will be discussed in more details later.

[6]It is assumed that $K \leq M$.

as

$$\mathcal{R}_{\text{TDMA}} \sim M \log\left(\frac{P}{M^2}\log N\right). \tag{98}$$

In this case, it can be easily shown that $\lim_{N\to\infty}\frac{\mathcal{R}_{\text{TDMA}}}{\mathcal{R}_{\text{Opt}}} = 1$. In other words, the optimum sum-rate can asymptotically be achieved. However, the selected dimensions by TDMA belong to the same user and have the asymptotic behavior of $\frac{\log N}{M}$, while in our proposed method the selected dimensions belong to different users with the asymptotic behvior of $\log N$. Moreover, we have

$$\begin{aligned}\mathcal{R}_{\text{Opt}} - \mathcal{R}_{\text{TDMA}} &\sim M\log\left(1+\frac{P}{M}\log N\right) - M\log\left(1+\frac{P}{M^2}\log N\right)\\ &\sim M\log M. \end{aligned} \tag{99}$$

As can be observed from figure 2, this gap affects the performance significantly, especially when $M$ is large.

*2) Random Selection:* In this method, the base station randomly selects $M$ users for transmission. This results in having fairness in the system. This strategy can also be regarded as Round-Robin scheduling algorithm, when the users are randomly divided into groups of size $M$, and the base station serves one group in each time slot.

In Appendix G, it is shown that using multiple dimensions for transmission results in having *multiplexing gain* equal to $M$. However, because of random selection of the users, this scheme does not provide multiuser diversity gain. More precisely,

$$\begin{aligned} d_{\mathrm{RS}} &= \lim_{N\to\infty} \frac{\mathcal{R}_{\mathrm{RS}}(P,N)}{M\log\log N} \\ &= \lim_{N\to\infty} \frac{\mathbb{E}_{\boldsymbol{H}_1,\cdots,\boldsymbol{H}_M}\left\{\max_{\substack{\boldsymbol{Q}_m \\ \sum \mathrm{Tr}(\boldsymbol{Q}_\mathrm{m})=\mathrm{P}}} \log\det\left(\boldsymbol{I}_M + \sum_{m=1}^{M} \boldsymbol{H}_m^* \boldsymbol{Q}_m \boldsymbol{H}_m\right)\right\}}{M\log\log N} \\ &= \lim_{N\to\infty} \frac{O(1)}{M\log\log N} \\ &= 0. \end{aligned} \tag{100}$$

As a result of lacking multiuser diversity gain, this scheme shows a weak performance especially for large number of users. (Figure 2)

### *C. Simulation Results*

So far, we have shown that as $N$ tends to infinity, our scheme achieves the optimum sum-rate which scales like $M\log\left(\frac{P}{M}\log N\right)$. In this section, simulation results are provided to examine the performance of our proposed scheme in practical networks with finite number of users. Figure 1 shows the optimum threshold (computed by exhaustive search) as a function of the number of users for $M=2, K=1$, and $M=4, K=1$. These curves show that the optimum threshold for each $N$, lies between $\log N - \log\log N$, and $\log N$.

Figures 2 presents the plots of the corresponding sum-rate versus the number of users for different number of transmit and receive antennas. The Signal to Noise Ratio (SNR), which is equal to the transmitted power $P$, is fixed to 10 dB in all curves. For comparison, the plots of sum-rate when using TDMA and Random Selection algorithms, as well as the optimum scheme of dirty-paper coding are also given. For Random Selection algorithm, it is assumed that the optimum precoding scheme of dirty-paper coding is used.

Figure 3 depicts the plots of sum-rate capacity versus SNR ($P$), for $M = 2, K = 1$ and $M = 4, K = 1$. The number of users is fixed to 100 in both curves. It can be observed that the sum-rate achieving by the proposed scheme shows a linear increase with $\log P$ in high SNRs with the slope equal to $M$. This confirms achieving the multiplexing gain of $M$ by the proposed scheme.

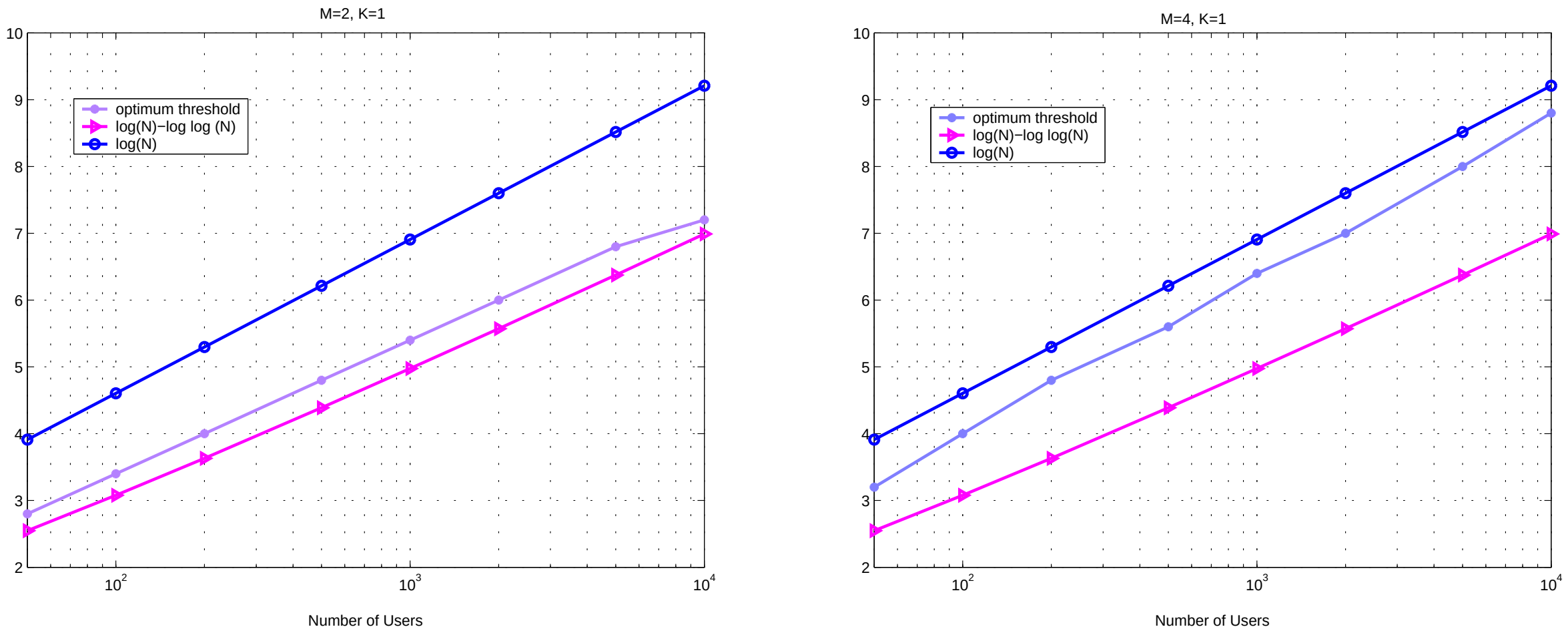

Fig. 1. Optimum threshold versus the number of users.

## V. Complexity Analysis

### A. Amount of Feedback

As can be observed in the proposed algorithm, only the eigenvectors that belong to $\mathcal{S}_0$, defined in (6), must be sent back to the base station, along with their corresponding singular values. For the asymptotic case of $N \to \infty$, from Lemma 2, we conclude that the cardinality of $\mathcal{S}_0$ scales as $\frac{e^{\rho(N)}}{\Gamma(M)\Gamma(K)}$. Assuming that for each eigenvector and its singular value $2M$ real values must be fed back, the total number of real values required at the base station is asymptotically equal to $\frac{2Me^{\rho(N)}}{\Gamma(M)\Gamma(K)}$.

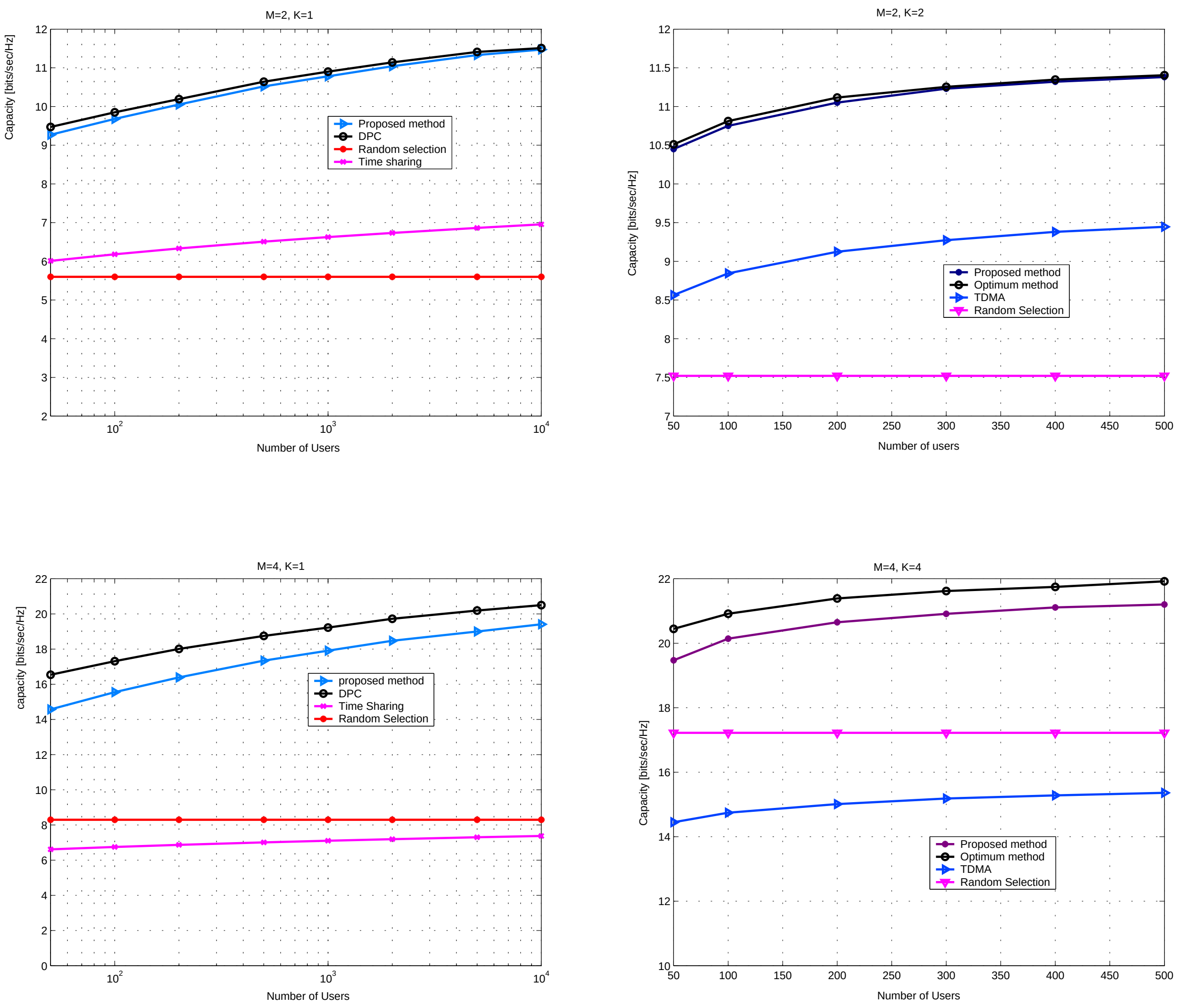

Fig. 2. Sum-rate capacity versus the number of users, $P = 10$dB.

From Theorem 1, we observe that to achieve the optimum sum-rate, i.e., $\lim_{N\to\infty} \mathcal{R}_{\text{Opt}} - \mathcal{R}_{\text{Prop}} = 0$, the following condition must be satisfied:

$$\rho(N) \sim \log\log\log\log N + \log[\Gamma(K)\Gamma(M)] + \omega\left(\frac{1}{\log\log\log N}\right). \tag{101}$$

As a result,

$$\mathcal{N}_{\text{Prop}} \sim 2M \log\log\log N + \omega(1), \tag{102}$$

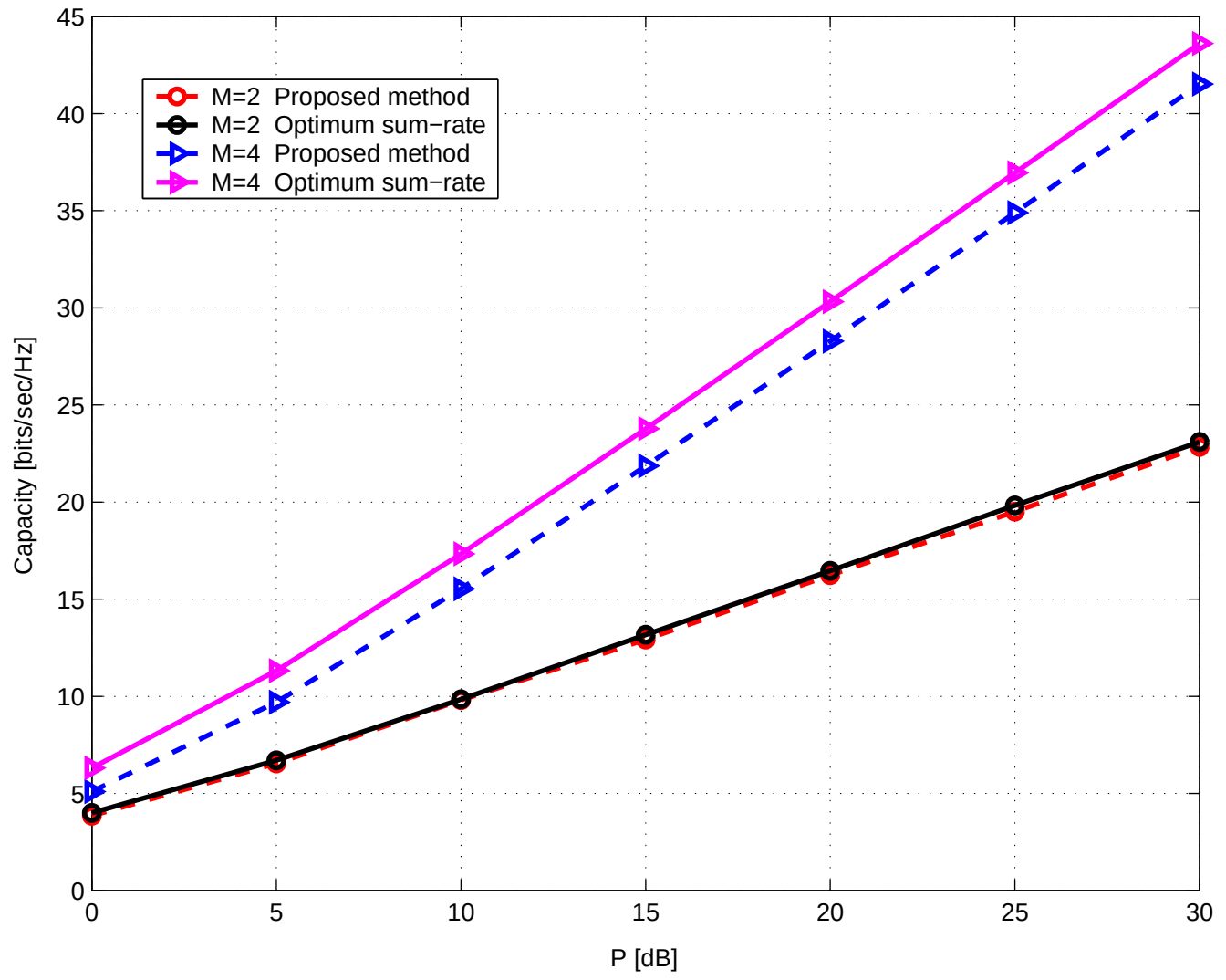

Fig. 3. Sum-rate capacity versus transmit power, $N = 100, K = 1$.

where $\mathcal{N}_{\text{Prop}}$ stands for the amount of feedback (in terms of the total number of real values required at the base station) in the proposed method. From the above equation, it follows that the minimum amount of feedback required to achieve the optimum performance is lower-bounded by $\log\log\log N$, in the proposed algorithm. However, in [28], it has been shown that the same result holds for any other strategies.

In order to guarantee $\lim_{N\to\infty} \mathcal{R}_{\text{Opt}} - \mathcal{R}_{\text{Prop}} = 0$ in the proposed scheme, using Theorem 2, the following condition must be satisfied:

$$\mathcal{N}_{\text{Prop}} \sim \omega(\log\log\log N). \tag{103}$$

Note that the computation of $\gamma_{k,j}$'s in Algorithm 1 (eq. (8)) can be performed in the mobile sides, which reduces the amount of feedback further. This idea is described in details as the following algorithm:

**Algorithm 2 (Modified version of Algorithm 1):**

1. Set the thresholds $t$ and $\beta$.
2. Define

$$\mathcal{S}_0 = \{(k,j)| \quad \lambda_j(k) > t\}.$$

   For all $(k,j) \in \mathcal{S}_0$, send $\lambda_j(k)$ to the base station.
3. Let $(s_1, d_1) = \arg \quad \max_{(k,j)\in\mathcal{S}_0} \lambda_j(k)$. Base station informs the user $s_1$ to feed back the eigenvector corresponding to its maximum singular value and after receiving it, sends these information to all the users in $\mathcal{S}_0 - \{(s_1, d_1)\}$.
4. Define $\gamma_{k,j}(0) = 0$ for all $(k,j) \in \mathcal{S}_0$. For $m = 1$ to $M-1$ the following steps are repeated:
   – Define $\mathcal{S}_m = \left\{(k,j)|(k,j) \in \mathcal{S}_{m-1}, |\boldsymbol{V}^*_{s_m,d_m}\boldsymbol{V}_{k,j}|^2 < \beta\right\}$ and $\gamma_{k,j}(m) = |\boldsymbol{V}^*_{s_m,d_m}\boldsymbol{V}_{k,j}|^2 + \gamma_{k,j}(m-1)$, for all $(k,j) \in \mathcal{S}_m$. All users in $\mathcal{S}_m$ feed back their corresponding $\gamma_{k,j}(m)$ to the base station.
   – Select $(s_{m+1}, d_{m+1}) = \arg \quad \min_{(k,j)\in\mathcal{S}_m} \quad \gamma_{k,j}(m)$. Base station inform the user $s_m$ to feedback its $d_m$th eigenvector, and after receiving, sends it to all users in $\mathcal{S}_m - -\{(s_m, d_m)\}$.

For the asymptotic case of $N \to \infty$, having $t = \log N + (M+K-2)\log\log N - \log\log\log\log N - q(N)$ and $\beta = e^{-\frac{q(N)}{M}}$, and using equations (53) and (64), we have

$$\begin{aligned}
\mathcal{N}_{\text{Prop}} &= \sum_{m=0}^{M-1} |\mathcal{S}_m| + 2M^2 \\
&\sim \sum_{m=0}^{M-1} L \times \text{Prob}\left\{k \in \mathcal{S}_m | k \in \mathcal{S}_0\right\} + 2M^2 \\
&\sim L + L\sum_{m=1}^{M-1} O(e^{-\frac{mq(N)}{M}}) + 2M^2 \\
&\sim L\left[1 + O\left(e^{-\frac{q(N)}{M}}\right)\right] \\
&\sim \frac{e^{\rho(N)}}{\Gamma(M)\Gamma(K)}.
\end{aligned} \tag{104}$$

Figure 4 depicts the plots of the required amount of feedback versus the number of users for $M = 2, K = 1$ and $M = 4, K = 1$, when Algorithm 1 and Algorithm 2 are used. The measure for the amount of feedback is defined as the number of real components per user that should be sent to the base station. In these curves, the optimum values for the thresholds ($t$ and $\beta$) are found by exhaustive search.

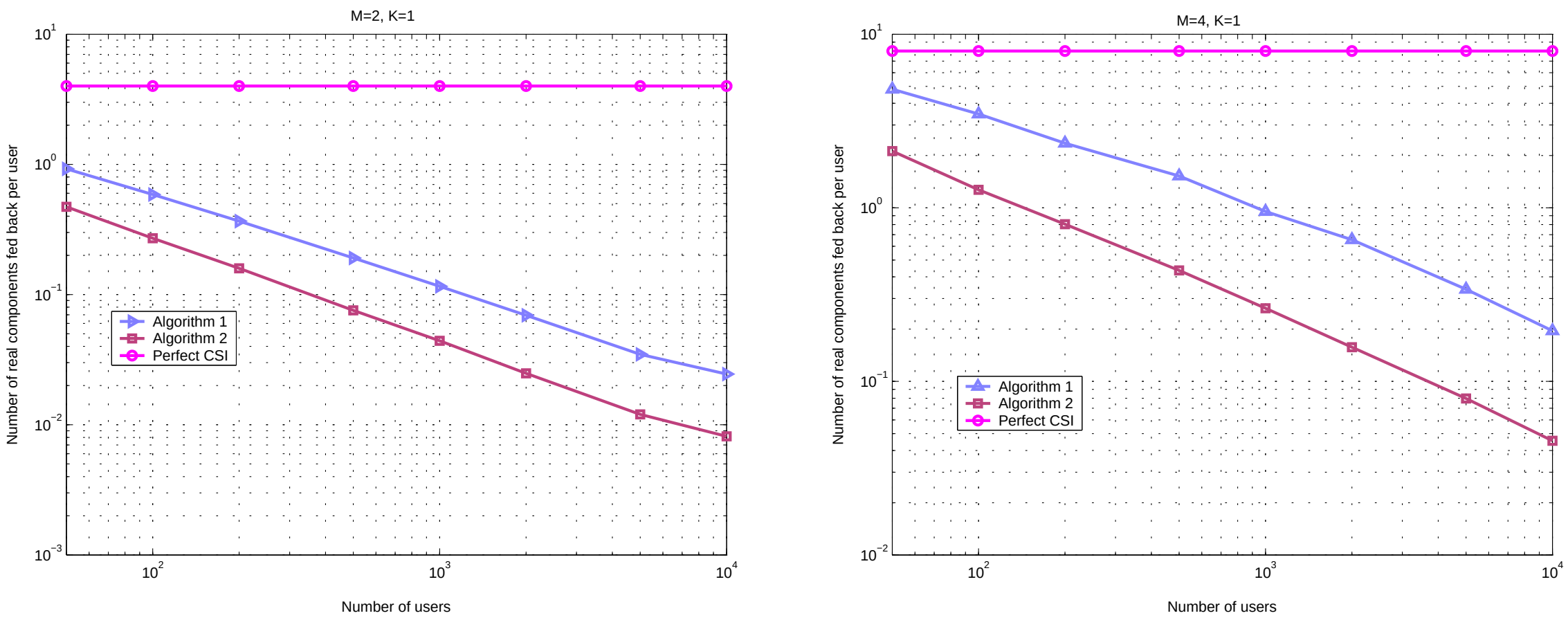

Fig. 4. Amount of feedback

## B. Search Complexity

Since at the first step of the algorithm, only a fraction of eigenvectors are pre-selected, the size of the search space for next steps is decreased from $NK$ to $L$. As can be observed, at the $m$th step of the algorithm, the base station searches for the dimension with the smallest $\gamma_{k,j}(m-1)$ among $\mathcal{S}_{m-1}$, which requires $L - m + 1$ searches. Therefore, the total number of searches for selecting the desired set is equal to $\sum_{m=1}^{M}(L - m + 1) = ML - \frac{M(M-1)}{2}$, which is linear in $L$. Again, we can restrict our search space if the modified algorithm stated in the previous section is used.

As mentioned earlier, the best $M$ eigenvectors for maximizing the sum-rate capacity can be found by exhaustive search. In this case, the size of the search space is equal to $\binom{NK}{M}$.

In the asymptotic case of $N \to \infty$, the total number of searches is $\Theta(e^{\rho(N)}) \sim o(N)$ for the proposed algorithm, which is much less than that of exhaustive search ($\Theta(N^M)$). Therefore, using our algorithm the complexity of search at the base station is decreased significantly.

## VI. Conclusion

In this paper, we have considered a downlink communication system, in which a base station equipped with $M$ transmit antennas communicates with $N$ users, each equipped with $K$ receive antennas. We have proposed an efficient suboptimum algorithm for selecting a set of users in order to maximize the sum-rate throughput of the system, using zero-forcing beam-forming at the base station. For the asymptotic case of $N \to \infty$, we have derived the necessary and sufficient conditions to achieve the optimum sum-rate capacity, such that $\lim_{N\to\infty} \mathcal{R}_{\text{Opt}} - \mathcal{R}_{\text{Prop}} = 0$. We have also investigated the complexity of our scheme in terms of the required amount of feedback from the users to the base station, as well as the number of searches needed for selecting the coordinates. The proposed algorithm is compared with some other downlink strategies like TDMA and Random Selection algorithms.

## Appendix A; Proof of Lemma 3

In this appendix, we derive the probability density function of $\mathcal{O}(i,j) = |\boldsymbol{V}^*_{i,\max}\boldsymbol{V}_{j,\max}|^2$. For simplicity of notation, $\boldsymbol{V}_{i,\max}$ is denoted by $\boldsymbol{\phi}_i$, and $\boldsymbol{V}_{j,\max}$ is denoted by $\boldsymbol{\phi}_j$. Since $\boldsymbol{\phi}_i$ and $\boldsymbol{\phi}_j$ are the eigenvectors of two independent matrices whose entries are independent $\mathcal{CN}(0,1)$, it follows from [34] that $\boldsymbol{\phi}_i$ and $\boldsymbol{\phi}_j$ are independent isotropically distributed unit vectors in $\mathbb{C}^M$,

with the following probability density function:

$$p_{\boldsymbol{\phi}_i}(\boldsymbol{\phi}) = p_{\boldsymbol{\phi}_j}(\boldsymbol{\phi}) = \frac{\Gamma(M)}{\pi^M}\delta(\boldsymbol{\phi}^*\boldsymbol{\phi} - 1). \tag{105}$$

Indeed, this probability density function does not change by multiplying any $M \times M$ unitary matrix $\boldsymbol{\Theta}$, i.e.,

$$p_{\boldsymbol{\Theta}\boldsymbol{\phi}_i}(\boldsymbol{\phi}) = p_{\boldsymbol{\phi}_i}(\boldsymbol{\phi}). \tag{106}$$

Now, define $u = \boldsymbol{\phi}_i^*\boldsymbol{\phi}_j$, and let $\boldsymbol{\Theta}$ be a unitary matrix whose first row is equal to $\boldsymbol{\phi}_i$. We can write

$$\begin{aligned} u &= \boldsymbol{\phi}_i^*\boldsymbol{\Theta}^*\boldsymbol{\Theta}\boldsymbol{\phi}_j \\ &= [\boldsymbol{\Theta}\boldsymbol{\phi}_i]^*\,\boldsymbol{\Theta}\boldsymbol{\phi}_j \\ &= [1\,0\,\cdots\,0]\,\boldsymbol{\phi}_j^{'} \\ &= \boldsymbol{\phi}_j^{'}(1), \end{aligned} \tag{107}$$

where $\boldsymbol{\phi}_j^{'} = \boldsymbol{\Theta}\boldsymbol{\phi}_j$, and $\boldsymbol{\phi}_j^{'}(1)$ is the first element of $\boldsymbol{\phi}_j^{'}$. Since $\boldsymbol{\Theta}$ is unitary, $\boldsymbol{\phi}_j$ and $\boldsymbol{\phi}_j^{'}$ have the same pdf. Hence, the probability density function of $\boldsymbol{\phi}_j^{'}(1)$ is the same as that of $\boldsymbol{\phi}_j(1)$, and can be computed as [34]

$$p_u(u) = p_{\boldsymbol{\phi}_j(1)}(u) = \frac{M-1}{\pi}\left(1 - |u|^2\right)^{M-2}. \tag{108}$$

Using the above equation, the probability density function of $\mathcal{O}(i,j) = |u|^2$ will be equal to

$$\begin{aligned} p_{\mathcal{O}(i,j)}(z) &= p_{|u|^2}(z) \\ &= \frac{p_{|u|}(\sqrt{z})}{2\sqrt{z}} \\ &= \frac{2\pi\sqrt{z}p_u(\sqrt{z})}{2\sqrt{z}} \\ &= (M-1)(1-z)^{M-2}. \end{aligned} \tag{109}$$

## APPENDIX B; PROOF OF (63)

Since the selected vectors $\left\{\boldsymbol{V}_{s_j,\max}\right\}_{j=1}^{i-1}$ are nearly orthogonal to each other, they form a basis for the sub-space spanned by them. We call this sub-space $\boldsymbol{\mathcal{P}}_{i-1}$. In the following, we denote $\boldsymbol{V}_{k,\max}$, the eigenvector corresponding to the maximum singular value of user $k$, by $\boldsymbol{\phi}_k$ for the simplicity of notation.

Any vector $\boldsymbol{v} \in \mathbb{C}^M$ can be represented as

$$\boldsymbol{v} \;=\; \boldsymbol{v}^{\perp} + \sum_{j=1}^{i-1} \left\langle \boldsymbol{\phi}_{s_j}, \boldsymbol{v} \right\rangle \boldsymbol{\phi}_{s_j}, \tag{110}$$

where $\boldsymbol{v}^{\perp}$ is the project of $\boldsymbol{v}$ on the null space of $\boldsymbol{\mathcal{P}}_{i-1}$, denoted by $\boldsymbol{\mathcal{P}}_{i-1}^{\perp}$, and $\left\langle \boldsymbol{\phi}_{s_j}, \boldsymbol{v} \right\rangle = \boldsymbol{\phi}_{s_j}^{*} \boldsymbol{v}$.

Defining the event $\mathcal{C}_i = \{\mathcal{O}(s_1,k) < \alpha, \cdots, \mathcal{O}(s_{i-1},k) < \alpha\}$ [7], the conditional probability in (63) can be written as

$$\kappa_i \;=\; \text{Prob}\left\{\mathcal{O}(s_i,k) < \alpha | \quad \mathcal{C}_i\right\}. \tag{111}$$

Using (54), we can write $\mathcal{C}_i$ by

$$\mathcal{C}_i \;=\; \left\{|\boldsymbol{\phi}_{s_1}^{*}\boldsymbol{\phi}_k|^2 < \alpha, \cdots, |\boldsymbol{\phi}_{s_{i-1}}^{*}\boldsymbol{\phi}_k|^2 < \alpha\right\}. \tag{112}$$

Hence, (111) can be expressed as

$$\kappa_i \;=\; \text{Prob}\left\{|\boldsymbol{\phi}_{s_i}^{*}\boldsymbol{\phi}_k|^2 < \alpha \;\middle|\; \quad |\boldsymbol{\phi}_{s_1}^{*}\boldsymbol{\phi}_k|^2 < \alpha, \cdots, |\boldsymbol{\phi}_{s_{i-1}}^{*}\boldsymbol{\phi}_k|^2 < \alpha\right\}. \tag{113}$$

Using (110), we can write $\boldsymbol{\phi}_k$ as

$$\boldsymbol{\phi}_k = \boldsymbol{\phi}_k^{\perp} + \sum_{j=1}^{i-1} \left\langle \boldsymbol{\phi}_{s_j}, \boldsymbol{\phi}_k \right\rangle \boldsymbol{\phi}_{s_j}, \tag{114}$$

and $\boldsymbol{\phi}_{s_i}$ as

$$\boldsymbol{\phi}_{s_i} = \boldsymbol{\phi}_{s_i}^{\perp} + \sum_{j=1}^{i-1} \left\langle \boldsymbol{\phi}_{s_j}, \boldsymbol{\phi}_{s_i} \right\rangle \boldsymbol{\phi}_{s_j}. \tag{115}$$

[7] Recall the definition of $\alpha$ which is $\frac{\epsilon(N)}{M}$.

Hence, $|\boldsymbol{\phi}_{s_i}^* \boldsymbol{\phi}_k|^2$ can be computed as,

$$\begin{aligned} |\boldsymbol{\phi}_{s_i}^* \boldsymbol{\phi}_k|^2 &= \Bigg| \left\langle \boldsymbol{\phi}_{s_i}^{\perp}, \boldsymbol{\phi}_k^{\perp} \right\rangle + \sum_{j=1}^{i-1} \left\langle \boldsymbol{\phi}_{s_i}, \boldsymbol{\phi}_{s_j} \right\rangle \left\langle \boldsymbol{\phi}_{s_j}, \boldsymbol{\phi}_k \right\rangle + \\ & \qquad \sum_{j=1}^{i-1} \sum_{\substack{l=1 \\ l \neq j}}^{i-1} \left\langle \boldsymbol{\phi}_{s_i}, \boldsymbol{\phi}_{s_j} \right\rangle \left\langle \boldsymbol{\phi}_{s_l}, \boldsymbol{\phi}_k \right\rangle \left\langle \boldsymbol{\phi}_{s_j}, \boldsymbol{\phi}_{s_l} \right\rangle \Bigg|^2 . \end{aligned} \tag{116}$$

Defining

$$\begin{aligned} u_1 &= \left\langle \boldsymbol{\phi}_{s_i}^{\perp}, \boldsymbol{\phi}_k^{\perp} \right\rangle, \\ u_2 &= \sum_{j=1}^{i-1} \left\langle \boldsymbol{\phi}_{s_i}, \boldsymbol{\phi}_{s_j} \right\rangle \left\langle \boldsymbol{\phi}_{s_j}, \boldsymbol{\phi}_k \right\rangle, \\ u_3 &= \sum_{j=1}^{i-1} \sum_{\substack{l=1 \\ l \neq j}}^{i-1} \left\langle \boldsymbol{\phi}_{s_i}, \boldsymbol{\phi}_{s_j} \right\rangle \left\langle \boldsymbol{\phi}_{s_l}, \boldsymbol{\phi}_k \right\rangle \left\langle \boldsymbol{\phi}_{s_j}, \boldsymbol{\phi}_{s_l} \right\rangle, \end{aligned} \tag{117}$$

we have

$$|\boldsymbol{\phi}_{s_i}^* \boldsymbol{\phi}_k|^2 = |u_1|^2 + |u_2|^2 + |u_3|^2 + 2\Re\{u_1 u_2^*\} + 2\Re\{u_2 u_3^*\} + 2\Re\{u_1 u_3^*\}, \tag{118}$$

where $\Re\{x\}$ denotes the real part of $x$. An upper bound for $|\boldsymbol{\phi}_{s_i}^* \boldsymbol{\phi}_k|^2$ is given by

$$|\boldsymbol{\phi}_{s_i}^* \boldsymbol{\phi}_k|^2 < |u_1|^2 + |u_2|^2 + |u_3|^2 + 2|u_1|(|u_2| + |u_3|) + 2|u_2||u_3|. \tag{119}$$

Having the facts that $\|\boldsymbol{\phi}_k^{\perp}\|^2 < \|\boldsymbol{\phi}_k\|^2 = 1$, and $\|\boldsymbol{\phi}_{s_i}^{\perp}\|^2 < \|\boldsymbol{\phi}_{s_i}\|^2 = 1$, we can write

$$\begin{aligned} |\boldsymbol{\phi}_{s_i}^* \boldsymbol{\phi}_k|^2 &< \frac{|u_1|^2}{\|\boldsymbol{\phi}_k^{\perp}\|^2 \|\boldsymbol{\phi}_{s_i}^{\perp}\|^2} + 2\frac{|u_1|}{\|\boldsymbol{\phi}_k^{\perp}\| \|\boldsymbol{\phi}_{s_i}^{\perp}\|}(|u_2| + |u_3|) + (|u_2| + |u_3|)^2 \\ &= \mathcal{O}\left(\boldsymbol{\phi}_k^{\perp}, \boldsymbol{\phi}_{s_i}^{\perp}\right) + 2\sqrt{\mathcal{O}\left(\boldsymbol{\phi}_k^{\perp}, \boldsymbol{\phi}_{s_i}^{\perp}\right)}(|u_2| + |u_3|) + (|u_2| + |u_3|)^2 \\ &= \left(\sqrt{\mathcal{O}\left(\boldsymbol{\phi}_k^{\perp}, \boldsymbol{\phi}_{s_i}^{\perp}\right)} + |u_2| + |u_3|\right)^2 . \end{aligned} \tag{120}$$

Also, a lower bound for $|\boldsymbol{\phi}_{s_i}^* \boldsymbol{\phi}_k|^2$ can be given as

$$\begin{aligned} |\boldsymbol{\phi}_{s_i}^* \boldsymbol{\phi}_k|^2 &> |u_1|^2 - 2|u_1|(|u_2| + |u_3|) - 2|u_2||u_3| \\ &> \mathcal{O}\left(\boldsymbol{\phi}_k^{\perp}, \boldsymbol{\phi}_{s_i}^{\perp}\right) \|\boldsymbol{\phi}_k^{\perp}\|^2 \|\boldsymbol{\phi}_{s_i}^{\perp}\|^2 - 2\sqrt{\mathcal{O}\left(\boldsymbol{\phi}_k^{\perp}, \boldsymbol{\phi}_{s_i}^{\perp}\right)}(|u_2| + |u_3|)\|\boldsymbol{\phi}_k^{\perp}\| \|\boldsymbol{\phi}_{s_i}^{\perp}\| - 2|u_2||u_3| \\ &> \mathcal{O}\left(\boldsymbol{\phi}_k^{\perp}, \boldsymbol{\phi}_{s_i}^{\perp}\right) \|\boldsymbol{\phi}_k^{\perp}\|^2 \|\boldsymbol{\phi}_{s_i}^{\perp}\|^2 - 2\sqrt{\mathcal{O}\left(\boldsymbol{\phi}_k^{\perp}, \boldsymbol{\phi}_{s_i}^{\perp}\right)}(|u_2| + |u_3|) - 2|u_2||u_3|. \end{aligned} \tag{121}$$

Using (114) and (115), we have

$$\|\boldsymbol{\phi}_k^{\perp}\|^2 = 1 - \sum_{j=1}^{i-1} |\boldsymbol{\phi}_{s_j}^* \boldsymbol{\phi}_k|^2 + \sum_{j=1}^{i-1} \sum_{\substack{l=1 \\ l \neq j}}^{i-1} \left\langle \boldsymbol{\phi}_k, \boldsymbol{\phi}_{s_j} \right\rangle \left\langle \boldsymbol{\phi}_{s_j}, \boldsymbol{\phi}_{s_l} \right\rangle \left\langle \boldsymbol{\phi}_{s_l}, \boldsymbol{\phi}_k \right\rangle, \tag{122}$$

and

$$\|\boldsymbol{\phi}_{s_i}^{\perp}\|^2 = 1 - \sum_{j=1}^{i-1} |\boldsymbol{\phi}_{s_j}^* \boldsymbol{\phi}_{s_i}|^2 + \sum_{j=1}^{i-1} \sum_{\substack{l=1 \\ l \neq j}}^{i-1} \left\langle \boldsymbol{\phi}_{s_i}, \boldsymbol{\phi}_{s_j} \right\rangle \left\langle \boldsymbol{\phi}_{s_j}, \boldsymbol{\phi}_{s_l} \right\rangle \left\langle \boldsymbol{\phi}_{s_l}, \boldsymbol{\phi}_{s_i} \right\rangle. \tag{123}$$

Conditioned on $\mathcal{C}_i$, and knowing that the set $\{\boldsymbol{\phi}_{s_j}\}_{j=1}^{i}$ is $\epsilon(N)$-orthogonal (or equivalently, $M\alpha$-orthogonal, i.e., $|\boldsymbol{\phi}_{s_j}^* \boldsymbol{\phi}_{s_l}|^2 < M\alpha, j, l = 1, \cdots, i$), from (117) we conclude the followings:

$$\begin{aligned} |u_2| &< (i-1)\sqrt{M}\alpha, \\ |u_3| &< (i-1)(i-2)M\alpha^{3/2}, \\ \|\boldsymbol{\phi}_k^{\perp}\|^2 &> 1 - (i-1)\alpha - (i-1)(i-2)\sqrt{M}\alpha^{3/2}, \\ \|\boldsymbol{\phi}_{s_i}^{\perp}\|^2 &> 1 - (i-1)M\alpha - (i-1)(i-2)M^{3/2}\alpha^{3/2}. \end{aligned} \tag{124}$$

Therefore, using (120), (121), and (124) the upper bound and lower bound for $|\boldsymbol{\phi}_{s_i}^* \boldsymbol{\phi}_k|^2$ can be rewritten as

$$|\boldsymbol{\phi}_{s_i}^* \boldsymbol{\phi}_k|^2 < \left( \sqrt{\mathcal{O}\left(\boldsymbol{\phi}_k^{\perp}, \boldsymbol{\phi}_{s_i}^{\perp}\right)} + (i-1)\sqrt{M}\alpha + (i-1)(i-2)M\alpha^{3/2} \right)^2, \tag{125}$$

and

$$|\boldsymbol{\phi}_{s_i}^* \boldsymbol{\phi}_k|^2 > A.\mathcal{O}\left(\boldsymbol{\phi}_k^{\perp}, \boldsymbol{\phi}_{s_i}^{\perp}\right) - 2B\sqrt{\mathcal{O}\left(\boldsymbol{\phi}_k^{\perp}, \boldsymbol{\phi}_{s_i}^{\perp}\right)} - C, \tag{126}$$

where $A = \left(1 - (i-1)\alpha - (i-1)(i-2)\sqrt{M}\alpha^{3/2}\right)\left(1 - (i-1)M\alpha - (i-1)(i-2)M\sqrt{M}\alpha^{3/2}\right)$, $B = (i-1)\sqrt{M}\alpha + (i-1)(i-2)M\alpha^{3/2}$, and $C = 2(i-1)^2(i-2)M^{3/2}\alpha^{5/2}$.

Using (111), (125), and (126) we have

$$\begin{aligned}
\kappa_i &> \text{Prob}\left\{\left[\sqrt{\mathcal{O}\left(\boldsymbol{\phi}_k^{\perp}, \boldsymbol{\phi}_{s_i}^{\perp}\right)} + (i-1)\sqrt{M}\alpha + (i-1)(i-2)M\alpha^{3/2}\right]^2 < \alpha\right\} \\
&= \text{Prob}\left\{\mathcal{O}\left(\boldsymbol{\phi}_k^{\perp}, \boldsymbol{\phi}_{s_i}^{\perp}\right) < \left[\sqrt{\alpha} - (i-1)\sqrt{M}\alpha + (i-1)(i-2)M\alpha^{3/2}\right]^2\right\} \\
&= \text{Prob}\left\{\mathcal{O}\left(\boldsymbol{\phi}_k^{\perp}, \boldsymbol{\phi}_{s_i}^{\perp}\right) < \alpha - 2(i-1)\sqrt{M}\alpha^{3/2} + O(\alpha^2)\right\}, \qquad (127)
\end{aligned}$$

and

$$\begin{aligned}
\kappa_i &< \text{Prob}\left\{A.\mathcal{O}\left(\boldsymbol{\phi}_k^{\perp}, \boldsymbol{\phi}_{s_i}^{\perp}\right) - 2B\sqrt{\mathcal{O}\left(\boldsymbol{\phi}_k^{\perp}, \boldsymbol{\phi}_{s_i}^{\perp}\right)} - C < \alpha\right\} \\
&= \text{Prob}\left\{\sqrt{\mathcal{O}\left(\boldsymbol{\phi}_k^{\perp}, \boldsymbol{\phi}_{s_i}^{\perp}\right)} < \frac{B + \sqrt{B^2 + A(C+\alpha)}}{A}\right\} \\
&= \text{Prob}\left\{\sqrt{\mathcal{O}\left(\boldsymbol{\phi}_k^{\perp}, \boldsymbol{\phi}_{s_i}^{\perp}\right)} < \sqrt{\alpha} + (i-1)\sqrt{M}\alpha + O(\alpha^{3/2})\right\} \\
&= \text{Prob}\left\{\mathcal{O}\left(\boldsymbol{\phi}_k^{\perp}, \boldsymbol{\phi}_{s_i}^{\perp}\right) < \alpha + 2(i-1)\sqrt{M}\alpha^{3/2} + O(\alpha^2)\right\}. \qquad (128)
\end{aligned}$$

Since $\boldsymbol{\phi}_k^{\perp}$ and $\boldsymbol{\phi}_{s_i}^{\perp}$ are the projections of $\boldsymbol{\phi}_k$ and $\boldsymbol{\phi}_{s_i}$ over $\boldsymbol{\mathcal{P}}_{i-1}^{\perp}$, a $(M-i+1)$-dimensional subspace of $\mathbb{C}^{M\times 1}$, $\frac{\boldsymbol{\phi}_k^{\perp}}{\|\boldsymbol{\phi}_k^{\perp}\|}$, and $\frac{\boldsymbol{\phi}_{s_i}^{\perp}}{\|\boldsymbol{\phi}_{s_i}^{\perp}\|}$, can be considered as uniformly distributed unit vectors in $\boldsymbol{\mathcal{P}}_{i-1}^{\perp}$. Therefore, using Lemma 3, the probability density function for $\mathcal{O}\left(\boldsymbol{\phi}_k^{\perp}, \boldsymbol{\phi}_{s_i}^{\perp}\right)$ can be given as

$$p_{\mathcal{O}\left(\boldsymbol{\phi}_k^{\perp}, \boldsymbol{\phi}_{s_i}^{\perp}\right)}(z) = (M-i)(1-z)^{M-i-1}. \qquad (129)$$

Having (129), and using (127) and (128) we can write

$$\begin{aligned}
\kappa_i &< \int_0^{\alpha + 2(i-1)\sqrt{M}\alpha^{3/2} + O(\alpha^2)} (M-i)(1-z)^{M-i-1} dz \\
&= 1 - \left[1 - \alpha - 2(i-1)\sqrt{M}\alpha^{3/2} + O(\alpha^2)\right]^{M-i} \\
&\sim (M-i)\alpha + 2(M-i)(i-1)\sqrt{M}\alpha^{3/2} + O(\alpha^2), \qquad (130)
\end{aligned}$$

and

$$\begin{aligned}\kappa_i &> \int_0^{\alpha-2(i-1)\sqrt{M}\alpha^{3/2}+O(\alpha^2)} (M-i)(1-z)^{M-i-1}dz \\ &= 1-\left[1-\alpha+2(i-1)\sqrt{M}\alpha^{3/2}+O(\alpha^2)\right]^{M-i} \\ &\sim (M-i)\alpha-2(M-i)(i-1)\sqrt{M}\alpha^{3/2}+O(\alpha^2). \end{aligned} \quad (131)$$

From (130) and (131) we conclude

$$\kappa_i \sim (M-i)\alpha + O(\alpha^{3/2}). \quad (132)$$

## APPENDIX C; PROOF OF LEMMA 5

Let us define

$$p = \text{Prob}\left\{\lambda_{\max}(\boldsymbol{H}_k) > t\right\}, \quad (133)$$

where $t = \log N + (M+K-1)\log\log N$. Using (25), the above probability probability can be written as

$$\begin{aligned} p &= \frac{t^{M+K-2}\exp(-t)}{\Gamma(M)\Gamma(K)}\left[1+O(t^{-1})\right] \\ &= \frac{[\log N + (M+K-1)\log\log N]^{M+K-2} + O\left([\log N]^{M+K-3}\right)}{\Gamma(M)\Gamma(K)e^{\log N+(M+K-1)\log\log N}} \\ &\sim \frac{1}{N\log N\Gamma(M)\Gamma(K)} + O\left(\frac{\log\log N}{N[\log N]^2}\right). \end{aligned} \quad (134)$$

Using the above equation, the probability in (75) can be computed as,

$$\begin{aligned} \eta &= 1-(1-p)^N \\ &\sim 1-\exp\left(-Np+O(Np^2)\right) \\ &\sim 1-\exp\left[-\frac{1}{\Gamma(M)\Gamma(K)\log N}+O\left(\frac{\log\log N}{[\log N]^2}\right)\right] \\ &\sim 1-\left[1-\frac{1}{\Gamma(M)\Gamma(K)\log N}+O\left(\frac{\log\log N}{[\log N]^2}\right)\right] \\ &\sim O\left(\frac{1}{\log N}\right). \end{aligned} \tag{135}$$

## Appendix D

We observed that $\boldsymbol{\mathcal{B}} = \boldsymbol{\mathcal{H}}\boldsymbol{\mathcal{H}}^*$ is an $M \times M$ matrix whose diagonal elements behave like $\log N + f(N)$, where $f(N) \sim o(\log N)$, and its non-diagonal elements scale as $O(\epsilon(N)\log N)$. For simplicity of notation, we define $\theta(N) = \log N + f(N)$ and $\varphi(N) = O(\epsilon(N)\log N)$.

Let us define $\mathcal{A}_m$ as a $m \times m$ matrix whose diagonal elements scale like $\theta(N)$, and, its non-diagonal elements scale like $\varphi(N)$. Hence, all diagonal elements of $\boldsymbol{\mathcal{B}}^{-1}$ can be written as $\frac{\det \mathcal{A}_{M-1}}{\det \mathcal{A}_M}$.

It can be easily shown that

$$\begin{aligned} \det \mathcal{A}_m &= [\theta(N)]^m + O([\theta(N)]^{m-2}[\varphi(N)]^2) \\ &= [\log N]^m + O\left([\log N]^m h(N)\right), \quad m = 2,\cdots,M. \end{aligned} \tag{136}$$

where $h(N) = \max\left(\frac{f(N)}{\log N}, \epsilon(N)\right) \sim o(1)$. Consequently, we can write any diagonal element of $\boldsymbol{\mathcal{B}}^{-1}$ as

$$\begin{aligned} [\boldsymbol{\mathcal{B}}^{-1}]_{ii} &= \frac{[\log N]^{M-1} + O\left([\log N]^{M-1}h(N)\right)}{[\log N]^M + O\left([\log N]^M h(N)\right)} \\ &= [\log N]^{-1} + O\left(h(N)[\log N]^{-1}\right). \end{aligned} \tag{137}$$

## APPENDIX E; PROOF OF LEMMA 6

For the proposed method, we have seen that the achievable sum-rate can be lower-bounded as

$$\begin{aligned} \mathcal{R}_{\text{Prop}} &\geq \mathbb{E}_{\boldsymbol{\mathcal{H}}}\left\{M\log\left(1+\frac{P}{\operatorname{Tr}\left\{[\boldsymbol{\mathcal{H}}^*\boldsymbol{\mathcal{H}}]^{-1}\right\}}\right)\right\} \\ &\geq M\log P - M\mathbb{E}_{\boldsymbol{\mathcal{H}}}\left\{\log\left(\operatorname{Tr}\left\{[\boldsymbol{\mathcal{H}}^*\boldsymbol{\mathcal{H}}]^{-1}\right\}\right)\right\}. \end{aligned} \tag{138}$$

where $\boldsymbol{\mathcal{H}}$ is the " selection coordinate matrix", defined in (9).

In [35], it has been shown that

$$\|\boldsymbol{b}_i\|^2\|\boldsymbol{a}_i\|^2 \leq \delta(\boldsymbol{B}), \qquad i = 1, \cdots, M, \tag{139}$$

where $\boldsymbol{b}_i$, $i = 1, \cdots, M$, are the columns of $\boldsymbol{B}$, a $M \times M$ matrix with the orthogonality defect $\delta(\boldsymbol{B})$, and $\boldsymbol{a}_i$, $i = 1, \cdots, M$, are the columns of $\boldsymbol{A} = (\boldsymbol{B}^{-1})^*$. Similarly, we can write

$$\|\boldsymbol{b}_i\|^2\|\boldsymbol{a}_i\|^2 \leq \delta(\boldsymbol{A}), \qquad i = 1, \cdots, M. \tag{140}$$

Defining $\boldsymbol{B} = \boldsymbol{\mathcal{H}}^{-1}$, and using the above equation, we can write

$$\begin{aligned} \operatorname{Tr}\left([\boldsymbol{\mathcal{H}}\boldsymbol{\mathcal{H}}^*]^{-1}\right) &= \sum_{i=1}^{M}\|\boldsymbol{b}_i\|^2 \\ &\leq \sum_{i=1}^{M}\frac{\delta(\boldsymbol{\mathcal{H}}^*)}{\|\boldsymbol{a}_i\|^2}, \end{aligned} \tag{141}$$

where $\boldsymbol{a}_i$, is the $i$th column of $\boldsymbol{\mathcal{H}}^*$, which is equal to $\boldsymbol{g}_{s_i}^*$. Having the fact that $\|\boldsymbol{g}_{s_i}\|^2 \geq t$ (by the algorithm), we can rewrite (141) as

$$\operatorname{Tr}\left([\boldsymbol{\mathcal{H}}\boldsymbol{\mathcal{H}}^*]^{-1}\right) \leq \frac{M\delta\left(\boldsymbol{\mathcal{H}}^*\right)}{t}. \tag{142}$$

Defining $X(\boldsymbol{\mathcal{H}}) = \log \operatorname{Tr}\left([\boldsymbol{\mathcal{H}}\boldsymbol{\mathcal{H}}^*]^{-1}\right)$, $Y(\boldsymbol{\mathcal{H}}) = \log \frac{M\delta\left(\boldsymbol{\mathcal{H}}^*\right)}{t}$, $Z(\boldsymbol{\mathcal{H}}) = \log \delta(\boldsymbol{\mathcal{H}}^*)$, and $F_W(.)$ as the CDF of the random variable $W$, we have

$$\begin{aligned} \mathbb{E}\left\{X(\boldsymbol{\mathcal{H}})\right\} &\leq \mathbb{E}\left\{Y(\boldsymbol{\mathcal{H}})\right\} \\ &= \log \frac{M}{t} + \mathbb{E}\{Z(\boldsymbol{\mathcal{H}})\} \\ &= \log \frac{M}{t} + \int_0^\infty z f_{Z(\boldsymbol{\mathcal{H}})}(z) dz \\ &= \log \frac{M}{t} + \int_0^\infty \left[1 - F_{Z(\boldsymbol{\mathcal{H}})}(z)\right] dz \\ &= \log \frac{M}{t} + \int_1^\infty \left[1 - F_{\delta(\boldsymbol{\mathcal{H}}^*)}(e^z)\right] dz. \end{aligned} \tag{143}$$

It can be easily shown that $\delta(\boldsymbol{\mathcal{H}}^*) = \delta(\boldsymbol{\Psi})$, where $\boldsymbol{\Psi} = [\boldsymbol{\Psi}_1|\cdots|\boldsymbol{\Psi}_M]$ is the matrix consisting of the normalized columns of $\boldsymbol{\mathcal{H}}^*$, i.e., $\boldsymbol{\Psi}_i = \frac{\boldsymbol{\mathcal{H}}^*{}_i}{\|\boldsymbol{\mathcal{H}}^*{}_i\|}$, $i = 1, \cdots, M$. Since the rows of $\boldsymbol{\mathcal{H}}$ are chosen randomly among the pre-selected eigenvectors, and due to the fact that the eigenvalues of a zero-mean circularly symmetric Gaussian matrix are independent of their corresponding eigenvectors, $\boldsymbol{\Psi}$ can be considered as a $M \times M$ matrix whose column are $M$ randomly selected unit vectors. We have

$$\begin{aligned} \delta(\boldsymbol{\Psi}) &= \frac{1}{|\det(\boldsymbol{\Psi})|^2} \\ &= \frac{1}{\prod_{i=1}^{M-1} \gamma_i}, \end{aligned} \tag{144}$$

where $\gamma_i$ is the square norm of the project of $\boldsymbol{\Psi}_{i+1}$ over the sub-space spanned by $\{\boldsymbol{\Psi}_j\}_{j=1}^i$, $\boldsymbol{\mathcal{P}}_i$. Now, consider $\boldsymbol{\Phi}_1, \cdots, \boldsymbol{\Phi}_M$, to be an orthonormal basis for the $M$-dimensional space, where $\{\boldsymbol{\Phi}_j\}_{j=1}^i$ are a basis for $\boldsymbol{\mathcal{P}}_i$. Therefore, $\boldsymbol{\Psi}_{i+1}$ can be represented as $(\psi_{1,i+1}, \cdots, \psi_{i,i+1}, 0, \cdots, 0)$, where $\psi_{j,i+1}$ is the project of $\boldsymbol{\Psi}_{i+1}$ over $\boldsymbol{\Phi}_j$. In [34], the joint probability density function of $\boldsymbol{\Psi}_{i+1}^{(i)} = (\psi_{1,i+1}, \cdots, \psi_{i,i+1})$ is given as,

$$p_{\boldsymbol{\Psi}_{i+1}^{(i)}}(\boldsymbol{\psi}) = \frac{\Gamma(M)}{\pi^i \Gamma(M-i)} \left(1 - \|\boldsymbol{\psi}\|^2\right)^{M-i-1}. \tag{145}$$

Using the above equation, the probability density function of $\gamma_i = \|\mathbf{\Psi}_{i+1}^{(i)}\|^2$ can be written as

$$p_{\gamma_i}(z) = \frac{\Gamma(M)}{\Gamma(i)\Gamma(M-i)} z^{i-1}(1-z)^{M-i-1}, \tag{146}$$

which corresponds to the Beta distribution with parameters $(i, M-i)$.

Using (144), (146), and independence of $\gamma_i$'s [19], we have

$$\begin{aligned} \text{Prob}\left\{\delta(\mathbf{\Psi}) > r\right\} &\leq \text{Prob}\left\{\min_i \gamma_i < r^{-\frac{1}{M-1}}\right\} \\ &= 1 - \prod_{i=1}^{M-1}\left[1 - I_{i,M-i}\left(r^{-\frac{1}{M-1}}\right)\right], \qquad r \geq 1, \end{aligned} \tag{147}$$

where $I_{r,s}(.)$ denotes the *Incomplete Beta Function*, with parameters $(r, s)$. In [36], it has been shown that

$$I_{r,s}(x) = \frac{\Gamma(r+s)x^r(1-x)^{s-1}}{\Gamma(r+1)\Gamma(s)} + I_{r+1,s-1}(x), \quad \forall r, s \in \mathbb{Z}^+, \tag{148}$$

which incurs that

$$I_{r,s}(x) \geq I_{r+1,s-1}(x), \quad \forall x \in [0,1]. \tag{149}$$

Consequently,

$$\begin{aligned} I_{i,M-i}(x) &\leq I_{1,M-1}(x) \\ &= 1 - (1-x)^{M-1}, \qquad i = 1, \cdots, M-1. \end{aligned} \tag{150}$$

Using (150) and (147), we can write,

$$\text{Prob}\left\{\delta(\mathbf{\Psi}) > r\right\} \leq 1 - \left(1 - \sqrt[M-1]{1/r}\right)^{(M-1)^2}. \tag{151}$$

Combining (143) and (151), we have

$$\begin{aligned}\mathbb{E}\{X(\boldsymbol{\mathcal{H}})\} &\leq \log\frac{M}{t}+\int_1^\infty\left[1-\left(1-e^{\frac{-r}{M-1}}\right)^{(M-1)^2}\right]dr \\ &= \log\frac{M}{t}+\sum_{m=1}^{(M-1)^2}\binom{(M-1)^2}{m}(-1)^{m+1}\int_1^\infty e^{-\frac{mr}{M-1}}dr \\ &= \log\frac{M}{t}+\sum_{m=1}^{(M-1)^2}\binom{(M-1)^2}{m}(-1)^{m+1}\frac{M-1}{m}e^{\frac{-m}{M-1}} \\ &= \log\frac{M}{t}+(M-1)\sum_{m=1}^{(M-1)^2}\frac{1-\left(1-e^{-\frac{1}{M-1}}\right)^m}{m} \\ &\leq \log\frac{M}{t}+(M-1)\sum_{m=1}^{(M-1)^2}\frac{1}{m} \\ &\leq \log\frac{M}{t}+(M-1)[2\log(M-1)+1]. \qquad (152)\end{aligned}$$

Substituting (152) into (138) and having $t=\log N$, we get

$$\mathcal{R}_{\text{Prop}}\geq M\log\left(\frac{P}{M}\log N\right)-M(M-1)[2\log(M-1)+1]. \qquad (153)$$

As a result,

$$\lim_{N\to\infty}\frac{\mathcal{R}_{\text{Prop}}}{\mathcal{R}_{\text{Opt}}}=1. \qquad (154)$$

## APPENDIX F; PROOF OF LEMMA 7

*Achievability of the maximum multiplexing gain*

Using (138), the multiplexing gain achieved by the proposed method, denoted by $r_{\text{Prop}}$, can be lower-bounded as

$$\begin{aligned}r_{\text{Prop}} &\geq \lim_{P\to\infty}\frac{M\log P-M\mathbb{E}_{\boldsymbol{\mathcal{H}}}\left\{\log\left(\text{Tr}\left\{[\boldsymbol{\mathcal{H}}^*\boldsymbol{\mathcal{H}}]^{-1}\right\}\right)\right\}}{\log P} \\ &= M-M\lim_{P\to\infty}\frac{\mathbb{E}_{\boldsymbol{\mathcal{H}}}\left\{\log\text{Tr}\left\{[\boldsymbol{\mathcal{H}}\boldsymbol{\mathcal{H}}^*]^{-1}\right\}\right\}}{\log P}. \qquad (155)\end{aligned}$$

Following the proof of Lemma 6 in Appendix E, and using equations (143), and (152), and the union bound for the probability, we have

$$\begin{aligned}\mathbb{E}_{\boldsymbol{\mathcal{H}}}\left\{\log \operatorname{Tr}\left\{[\boldsymbol{\mathcal{H}}\boldsymbol{\mathcal{H}}^*]^{-1}\right\}\right\} &\leq \log\frac{M}{t} + \binom{L}{M}\int_1^{\infty}\left[1-\left(1-e^{\frac{-r}{M-1}}\right)^{(M-1)^2}\right]dr \\ &\leq \log\frac{M}{t} + (M-1)[2\log(M-1)+1]\binom{L}{M}, \qquad (156)\end{aligned}$$

where $L$ is the number of preselected eigenvectors in the first step of Algorithm 1. Since $L \leq NK$, we have $\mathbb{E}_{\boldsymbol{\mathcal{H}}}\left\{\log \operatorname{Tr}\left\{[\boldsymbol{\mathcal{H}}\boldsymbol{\mathcal{H}}^*]^{-1}\right\}\right\} < \infty$, the second term in (155) approaches zero, and as a result $r_{\text{Prop}} \geq M$.

For the optimum strategy, the sum-rate can be upper-bounded as [37],

$$\mathcal{R}_{\text{Opt}} \leq M\mathbb{E}_{\|\boldsymbol{H}\|_{\max}}\left\{\log\left(1+\frac{P}{M}\|\boldsymbol{H}\|^2_{\max}\right)\right\}, \qquad (157)$$

where $\|\boldsymbol{H}\|^2_{\max}$ is the maximum Frobinous norm of all channel matrices. This random variable can be considered as the maximum of $N$ $\chi^2(2MK)$ random variables which has the pdf of the form

$$p_{\|\boldsymbol{H}\|^2_{\max}}(x) = N\frac{x^{MK-1}\exp(-x)}{\Gamma(MK)}\gamma(x,MK)^{N-1}, \qquad (158)$$

where $\gamma(x,MK) = \int_x^{\infty}\frac{u^{MK}\exp(-u)}{\Gamma(MK)}du$. So, using (157) and (158), we can write the upper bound for the sum-rate as

$$\mathcal{R}_{\text{Opt}} \leq M\int_0^{\infty}\log(1+\frac{P}{M}x)N\frac{x^{MK-1}\exp(-x)}{\Gamma(MK)}\gamma(x,MK)^{N-1}dx. \qquad (159)$$

Thus, using the above equation, we have

$$\begin{aligned} r_{\text{Opt}} &= \lim_{P\to\infty}\frac{\mathcal{R}_{\text{Opt}}}{\log P} \\ &\leq \frac{M\log P + \int_0^{\infty} M\log(\frac{x}{M})N\frac{x^{MK-1}\exp(-x)}{\Gamma(MK)}\gamma(x,MK)^{N-1}dx}{\log P} \\ &= M. \qquad (160)\end{aligned}$$

Since for any values of $P$ and $N$, $\mathcal{R}_{\text{Opt}}(P, N)$ is the maximum achievable sum-rate , $r_{\text{Opt}}$ will be the maximum achievable multiplexing gain in MIMO-BC. Hence, using the above equation and having the fact that $r_{\text{Prop}} \geq M$, we conclude $r_{\text{Opt}} = r_{\text{Prop}} = M$ Therefore, the proposed method achieves the maximum *multiplexing gain* in MIMO-BC.

*Achievability of the optimum multiuser diversity gain*

In the proof of theorem 2, we observed that the sum-rate achieved by the proposed strategy, as well as the optimum one, scales like $M \log\left(\frac{P}{M}\log N\right)$. Hence, using (162) the *multiuser diversity gain* for the optimal scheme, denoted by $d_{\text{Opt}}$ is equal to

$$\begin{aligned} d_{\text{Opt}} &= \lim_{N\to\infty} \frac{\mathcal{R}_{\text{Opt}}}{r_{\text{Opt}} \log\log N} \\ &= \lim_{N\to\infty} \frac{M \log\left(\dfrac{P}{M}\log N\right)}{M \log\log N} \\ &= 1. \end{aligned} \tag{161}$$

and for the proposed method,

$$\begin{aligned} d_{\text{Prop}} &= \lim_{N\to\infty} \frac{\mathcal{R}_{\text{Prop}}}{r_{\text{Prop}} \log\log N} \\ &= \lim_{N\to\infty} \frac{M \log\left(\dfrac{P}{M}\log N\right)}{M \log\log N} \\ &= 1. \end{aligned} \tag{162}$$

Therefore, the proposed method achieves the maximum *multiuser diversity gain* in MIMO-BC. This, completes the proof of Lemma 7.

## Appendix G; Multiplexing Gain in Random Selection Method

In this appendix, we prove that the Random selection strategy achieves the maximum *multiplexing gain*, i.e., $r_{\text{RS}} = M$. For this purpose, we consider the the precoding scheme of zero-forcing beam-forming. We assume that the coordinates are chosen randomly among the eigenvectors corresponding to the maximum singular value of each user's channel matrix. Therefore, similar to (155), we have

$$r_{\mathrm{RS}}^{\mathrm{ZFBF}} \geq M - M \lim_{P\to\infty} \frac{\mathbb{E}_{\boldsymbol{\mathcal{H}}}\left\{\log \mathrm{Tr}\left\{[\boldsymbol{\mathcal{H}}^*\boldsymbol{\mathcal{H}}]^{-1}\right\}\right\}}{\log P}, \tag{163}$$

where $\boldsymbol{\mathcal{H}} = \left[\boldsymbol{g}_{s_1,\max}^T \,\middle|\, \boldsymbol{g}_{s_2,\max}^T \,\middle|\, \cdots \,\middle|\, \boldsymbol{g}_{s_M,\max}^T\right]^T$, and the users $s_1, \cdots, s_M$ are selected randomly. Defining $\boldsymbol{B} = \boldsymbol{\mathcal{H}}^{-1}$, similar to (141), we can write

$$\mathrm{Tr}\left\{[\boldsymbol{\mathcal{H}}\boldsymbol{\mathcal{H}}^*]^{-1}\right\} \leq \sum_{i=1}^{M} \frac{\delta(\boldsymbol{\mathcal{H}}^*)}{\|\boldsymbol{a}_i\|^2}, \tag{164}$$

where $\boldsymbol{a}_i$ is the $i$th column of $\boldsymbol{\mathcal{H}}^*$, which is equal to $\boldsymbol{g}_{s_i}$. Noting that $\|\boldsymbol{g}_{s_i}\|^2 = \lambda_{\max}(\boldsymbol{H}_{s_i})$, we have

$$\begin{aligned} \mathrm{Tr}\left\{[\boldsymbol{\mathcal{H}}\boldsymbol{\mathcal{H}}^*]^{-1}\right\} &\leq \sum_{i=1}^{M} \frac{\delta(\boldsymbol{\mathcal{H}}^*)}{\lambda_{\max}(\boldsymbol{H}_{s_i})} \\ &\leq \sum_{i=1}^{M} \frac{M\delta(\boldsymbol{\mathcal{H}}^*)}{\|\boldsymbol{H}_{s_i}\|^2}. \end{aligned} \tag{165}$$

Using (143), (152), and (165) we can write

$$\begin{aligned} \mathbb{E}_{\boldsymbol{\mathcal{H}}}\left\{\log \mathrm{Tr}\left\{[\boldsymbol{\mathcal{H}}^*\boldsymbol{\mathcal{H}}]^{-1}\right\}\right\} &\leq \mathbb{E}\left\{\log\left(\sum_{i=1}^{M} \frac{M\delta(\boldsymbol{\mathcal{H}}^*)}{\|\boldsymbol{H}_{s_i}\|^2}\right)\right\} \\ &= \log M + \mathbb{E}\left\{\log \delta(\boldsymbol{\mathcal{H}}^*)\right\} + \mathbb{E}\left\{\log\left(\sum_{i=1}^{M} \frac{1}{\|\boldsymbol{H}_{s_i}\|^2}\right)\right\} \\ &\leq \log M + (M-1)[2\log(M-1)+1] + \log\left[M\mathbb{E}\left\{\frac{1}{\|\boldsymbol{H}_{s_i}\|^2}\right\}\right] \\ &\leq M[2\log(M-1)+1] + \log\left[\int_0^\infty x^{-1}.\frac{x^{MK-1}\exp(-x)}{\Gamma(MK)}dx\right] \\ &= M[2\log(M-1)+1] - \log(MK-1). \end{aligned} \tag{166}$$

Using (163) and (166), and noting that $r_{\mathrm{RS}}^{\mathrm{ZFBF}} \leq r_{\mathrm{RS}} \leq M$, we conclude $r_{\mathrm{RS}} = M$.